\documentclass[pre, twocolumn,
 amssymb,
 aps]{revtex4-1}

 \usepackage{graphicx}
 \usepackage[cmex10]{amsmath}
 
\usepackage{mathtools, amssymb, graphicx, amsfonts}
\usepackage{xcolor, comment} 
\usepackage{bbm} 
\usepackage{float} 
\usepackage{amsopn} 
\usepackage{mathrsfs} 
\usepackage[ruled,vlined]{algorithm2e}
\usepackage[colorlinks=true, urlcolor=blue, citecolor=black]{hyperref}

\usepackage{bm}

\begin{document}

\title{Strong and Weak Random Walks on Signed Networks}

\author{S. A. Babul\textsuperscript{1,2,*}, Y. Tian\textsuperscript{3,†} and
R. Lambiotte\textsuperscript{1,2,‡}}
\affiliation{%
1. Mathematical Institute, University of Oxford, Woodstock Road, Oxford, OX2 6GG, United Kingdom
\\
2. The Turing Institute, British Library, 96 Euston Rd, London, NW1 2DB, United Kingdom\\
3. Nordita, Stockholm University and KTH Royal Institute of Technology, Sweden
\\
{\normalfont{\textsuperscript{*}shazia.babul@new.ox.ac.uk, 
\textsuperscript{†}yu.tian@su.se, \textsuperscript{‡}lambiotte@maths.ox.ac.uk}
}}

\begin{abstract}
Random walks play an important role in probing the structure of complex networks. On traditional networks, they can be used to extract community structure, understand node centrality, perform link prediction, or capture the similarity between nodes. On signed networks, where the edge weights can be either positive or negative, it is non-trivial to design a random walk which can be used to extract information about the signed structure of the network, in particular the ability to partition the graph into communities with positive edges inside and negative edges in between. Prior works on signed network random walks focus on the case where there are only two such communities (strong balance), which is rarely the case in empirical networks. In this paper, we propose a signed network random walk which can capture the structure of a network with more than two such communities (weak balance). The walk results in a similarity matrix which can be used to cluster the nodes into antagonistic communities. We compare the characteristics of the so-called strong and weak random walks, in terms of walk length and stationarity. We show through a series of experiments on synthetic and empirical networks that the similarity matrix based on weak walks can be used for both unsupervised and semi-supervised clustering, outperforming the same similarity matrix based on strong  walks when the graph has more than two communities, or exhibits asymmetry in the density of links. These results suggest that other random-walk based algorithms for signed networks could be improved simply by running them with weak walks instead of strong walks. 

\end{abstract}

\maketitle

\section{Introduction}

Random walks play a central role in several aspects of network science \cite{Masuda2017RW}. Random walks are often used as a linear, tractable model for diffusive processes on networks, allowing us to explore questions such as the impact of certain network properties \cite{lambiotte2021modularity}, e.g.~their community structure or their degree distribution, on the speed of the diffusion, e.g.~via the mixing time \cite{levin2017markov}. Random walks are also a key ingredient in several methods to extract information from large-scale networks. Important areas of application include node centrality which can be captured by the density of walkers on that node in the stationary state \cite{gleich2015pagerank}, community detection where groups of nodes can be defined as communities based on their tendency to retain flows of walkers for long times \cite{rosvall2008maps,lambiotte2014random}, or the design of kernels/embeddings capturing the similarity between nodes on a graph \cite{fouss2016algorithms,grover2016node2vec}.
Among the advantages of random-walk based approaches, one can cite their intuitive interpretation, in terms of flows along the edges, their connection to well-established theories, such as that of Markov chains, their linearity, opening up the toolbox of linear algebra, and their ease of use in a distributed setting, each of the walkers being independent from the others, and their capacity to probe the structure of the network at multiple scales.

In this article, our purpose is to explore how random walks can be defined on, and help us understand the structure of, signed networks. In signed networks, edges carry a value, that is positive or negative to model  trustful or mistrustful, activatory or inhibitory, cooperative or antagonistic interactions, depending on the context \cite{altafini2012dynamics,traag2019partitioning,szell2010multirelational}. Random walks on signed networks have been studied in the past, either implicitly, by investigating the properties of the signed Laplacian e.g.~for node embedding \cite{kunegis2010spectral} or explicitly, by defining rules for how traversing a negative edge affects the random walker \cite{tian2024spreading}, in a model that we will refer to as a strong random walk for reasons that will appear clear later. 
As in many generalisation of ``classical networks", generalising the notion of random walks to signed networks is not unique, and the choice should be motivated either by the process that one aims to model or by the underlying structure that one would like to extract. What makes the power of random walks on unsigned networks is their deep connection to specific structural patterns,  communities and bottlenecks, e.g.~via the Cheeger inequality \cite{cheeger1970lower}. Communities can be seen as noisy connected components, i.e. connected components to which random edges have been added. The important question to define the {\it right} model of random walks on signed networks is thus: what is the prototypical structure that one aims to identify? 

The large-scale structure of signed networks is usually viewed through the prism of balance theory. There exist two forms of balance. Strong balance \cite{harary1953notion} refers to the situation when the network can be partitioned into two groups, such that positive edges are inside the groups, and negative edges between them. Equivalently, balanced networks are such that any cycle in the graph has an even number of negative edges. 
A weaker version of balance \cite{davis1967clustering} relies on the condition that there is no cycle with only one single negative edge. Graphs that exhibit weak balance can be partitioned into $k$ clusters with positive edges inside, and negative edges connecting them. It is well-known that the signed Laplacian and the equivalent strong random walk allow to efficiently recover networks with polarisation between two groups, akin to strong balance \cite{kunegis2010spectral}. It is also well-known that they may fail to uncover and properly characterise networks exhibiting a structure related to weak balance, and thus to serve as a model for the inference of a structure into more than two groups \cite{traag2019partitioning}. Alternative approaches have been proposed to solve this problem, for instance through other notions of Laplacian, e.g. SPONGE \cite{cucuringu2019sponge} or the repelling Laplacian \cite{babul2024sheep}, but they do not have a random walk interpretation and thus do not inherit their advantages described above. The main contribution of this article
is the definition of a random walk process, called the weak random walk, specifically designed for structures derived from weak balance, hence particularly fitted for clustering problems involving more than two groups in signed networks. In essence, the process consists in allowing the walker to explore the graph around a seed node only until it encounters a negative edge for a second time. 

This article is organised as follows. In section II, we provide an overview of the key concepts on unsigned networks and random walks, in particular related to the definition of a similarity matrix between nodes. In section III, we introduce a first type of random walk on signed networks, called the strong walk, and show its connections to the signed Laplacian. In section IV, we then discuss how to exploit such a random walk in order to define a similarity matrix between nodes of a signed network. In section V, we identify the limitations of the strong walks for networks exhibiting more than two clusters, and introduce weak walks to remedy the problem. 
 In section VI and VII, we compare the results of simple unsupervised and semi-supervised algorithms based on the two types of walks, showing that weak walks present an advantage in terms of accuracy for certain families of networks. Finally, in section VIII, we conclude and discuss implications of our work for the design of algorithms.

\section{Notations and basics}
We consider an undirected signed graph, characterised by the adjacency matrix $A_{s;ij}$, so that $A_{s;ij}=1$ indicates the presence of a positive edge. $A_{s;ij}=-1$ indicates the presence of a negative edge. We decompose that matrix into two matrices, encoding the positive and negative edges separately, each associated to an unsigned, undirected network: $A_{s;ij} = A_{+;ij} - A_{-;ij}$. In the signed graph, the total number of connections of each node, positive or negative, is thus given by $k_i = \sum_j |A_{s;ij}|$, which defines the degree. The total number of nodes is denoted as $N$, and that of edges is denoted by $M$. In the following, we will also use the diagonal matrix of degrees $K_{ij}=k_i \delta_{ij}$ and the adjacency matrix of the graph where signs are not considered $A_{u;ij} = A_{+;ij} + A_{-;ij}$. Note that $k_i = \sum_j A_{u;ij}$.
Our purpose is to model diffusive processes on signed networks. Here, we will focus on discrete-time walks, but everything can be generalised to continuous times; we consider unweighted networks, but the generalisation to weighted ones is straightforward. 

In the case when the graph does not have negative edges $A_{u;ij} = A_{+;ij}$, a discrete-time random walk  evolves according to 
\begin{eqnarray}
n_j(t+1) = \sum_i n_i(t) T_{u;ij}
\end{eqnarray}
where $n_i(t)$ is the density of walkers on node $i$ at time $t$ and $T_u =  K^{-1} A_u$ is the transition matrix of the associated Markov chain. This equation describes a microscopic process where an ensemble of walkers all move synchronously at discrete times. At each step, a walker located on some node $i$ jumps to one of its neighbours, with a probability equal to one over the number of such neighbouring nodes. The spectral properties of the transition matrix, and how they relate to the associated graph structure, are well known \cite{Masuda2017RW}. In particular, we note that $T_u$ shares the same spectrum as the symmetric matrix $K^{-1/2} A_u K^{-1/2}$, which implies that its eigenvalues are all real and that its left and right eigenvectors, ${\bf u}_{\alpha}$ and ${\bf v}_{\alpha}$ respectively, are related by ${\bf u}_{\alpha} = K  {\bf v}_{\alpha}$. The eigenvalues $\lambda_\alpha$ are in the interval $[-1,1]$, and the multiplicity of the dominant eigenvalue, $\lambda_1=1$ is equal to the number of connected components in the graph. The difference between $\lambda_1$ and the largest non-one eigenvalue, $\lambda_1-\lambda_2$, called the spectral gap, determines the mixing time of the process and quantifies the presence of bottlenecks hindering the flow of walkers across the system \cite{lambiotte2021modularity}. When the graph is connected, its corresponding left eigenvector, denoted by ${\bf \pi}$ is given by  $\pi_i=k_i/(2M)$, which characterises how walkers are distributed across the nodes in the stationary state.

As we mentioned in the introduction, random walks are at the core of several algorithms \cite{lambiotte2014random,perozzi2014deepwalk}. In particular, they are useful to capture the strength of indirect connections between nodes. 
In situations when the graph is sparse, so that only a small fraction of the pairs of nodes are connected \footnote{This is the case in the vast majority of real-world applications, especially when the graph is sufficiently large.}, random-walk-based kernels allow to estimate the proximity of pairs of nodes that are not directly connected, by incorporating, in an unbiased (or in a supervised) way, the information of walks of any length. In case of unsigned networks, the resulting similarity between two nodes is obtained from an appropriately weighted sum of the walks between them. Typically, the existence of many short walks between two nodes ensures their proximity. Important examples include the diffusion, or heat kernels \cite{kondor2002DiffusionKO,lawler2010heat,smola2003heat}, the mean commute time \cite{fouss2007communte,saerens2004commute}, but also the random-walk with restart similarity \cite{pan2004rwr,tong2007rwr,tong2008rwr}. For reasons that will appear clear later, and that random walks in our model on signed networks tend to stop after a finite number of steps, the latter generalises most naturally (according to us) in the signed context. 
 
Random walks with restart are a special type of random walk with teleportation. Take a node $i$, we consider a process where a walker may, at each step, either perform a jump to one of the neighbouring nodes with probability $p$, or teleport to the original node $i$ with probability $(1-p)$. It is straightforward to show that its stationary state is given by
 \begin{eqnarray}
{\bf n} = (1-p) {\bf e}_i (I - p T_u)^{-1},
\end{eqnarray}
where ${\bf e}_i$ is a row vector made of zeros except for entry $i$ where it takes the value $1$. The stationary state is, intuitively, peaked around the pinned node $i$, and becomes more spread, giving more importance to distant neighbours, when $p$ increases. When $p$ approaches $1$, more and more importance is given to longer walks, and the vector converges to the stationary state $\pi$ defined above. After taking a Taylor series of this expression, it is very similar to the kernel used in Deepwalk \cite{qiu2018network}, 
and defines the weighted neighborhood of each node at different scales. 

A similarity matrix can be obtained by repeating the process for each node and by stacking the $N$ vectors, leading to
 \begin{eqnarray}
 \label{kernel1}
\kappa(p) = (1-p) K (I - p T_u)^{-1},
\end{eqnarray}
 where we scale here each line by the node degree, to ensure that the matrix is symmetric. This quantity, or variations around it, have been used in a variety of contexts~\cite{fouss2016algorithms}, including node classification and clustering.
Note that
\begin{eqnarray}
\kappa_{ij}(p) &=& (1-p) k_i [(I - p T_u)^{-1}]_{ij}, \cr
\kappa_{ij}(p) &=& (1-p) k_i [\sum_k p^k T_u^{k}]_{ij}.
\end{eqnarray}
As each term $k_i  T^{k}_{u;ij}$ takes the $(i,j)$ element from $A_u K^{-1} A_u  .... K^{-1} A_u$, $\kappa(p)$ is symmetric for any value of $p$ \footnote{The symmetry arises due to the detailed balance when teleportation to nodes is done proportional to the stationary state of the walk without restart.}. Moreover, the $i^{\rm th}$ line (row) sums to $k_i$, meaning that $\kappa(p)$ can be interpreted as a weighted graph, interpolating between the original graph (with the addition of self-loops) when $p$ approaches $0$ and low rank matrix obtained from the outer product of the stationary states, $k_ik_j/(2M)$, when $p$ approaches $1$. For any value of $p$, $\kappa_{ij}(p)$ can be fed into standard methods for weighted networks. 
For instance, its Newman-Girvan modularity \cite{girvan2002modularity} can be used to uncover its multi-scale modularity, an approach reminiscent of Markov stability \cite{delvenne2010markovs,lambiotte2014rws,schaub2012rws}.

\section{Strong walks}

\subsection{Introduction and basic properties}

To generalize the classical methods defined in the previous section and, more generally, any algorithm based on random walks, one needs ways to define a random walk on a signed network. 
The natural way to do it is to consider the dynamics of two types of walkers, positive walkers and negative walkers \cite{tian2024spreading}. In the following, we will therefore start from describing how the number of positive walkers, $n_{+;i}(t)$, and that of negative walkers, $n_{-;i}(t)$, on each node $i$ evolve in time, so as to define different versions of random walks on signed networks. In this section, we give an interpretation of established methods for signed networks based on the signed Laplacian in the language of random walks.

The strong random walk, as we call it, is defined as follows. Positive and negative walkers perform a random walk by taking an edge, independently of its sign. If the walker takes a positive edge, it keeps its own sign; else it changes sign. The rate equation for the process is: 

\begin{eqnarray}
\label{discrete1}
n_{+;j}(t+1) = \sum_i \left( n_{+;i} T_{+;ij}  + n_{-;i} T_{-;ij}  \right) , \cr
n_{-;j}(t+1) = \sum_i \left( n_{-;i} T_{+;ij}  +  n_{+;i} T_{-;ij}  \right),
\end{eqnarray}
where 
\begin{eqnarray}
T_{+;ij} =  A_{+;ij}/k_i, \cr
T_{-;ij} =  A_{-;ij}/k_i,
\end{eqnarray}
indicate the transition probabilities for a walker from $i$ to $j$ and whether to keep or change the sign respectively. In matrix notation:
\begin{eqnarray}
T_+ =  K^{-1} A_+, \cr
T_- =  K^{-1} A_-.
\end{eqnarray}

If we consider the dynamics of the walkers independently of their sign and consider the time evolution of $n_i(t) = n_{+;i}(t) + n_{-;i}(t)$, one recovers the classical equation for random walk on networks
\begin{eqnarray}
n_j(t+1) = \sum_i  n_i(t) T_{u;ij},
\end{eqnarray}
where $T_u =  K^{-1} A_u$. If, in contrast, we consider the time evolution of the  difference $\Delta_i = n_{+;i} - n_{-;i}$, that can be understood as the total ``charge" on node $i$, we find 
\begin{eqnarray}
 \Delta_j(t+1) = \sum_i  \Delta_i(t) T_{s;ij},
\end{eqnarray}
where $T_s =  K^{-1} A_s,$ is the signed transition matrix from which one can build the standard signed  Laplacian, whose spectral properties are closely associated to the notion of strong balance and have been used for node embeddings and clustering \cite{kunegis2010spectral,traag2019partitioning}.

Another interesting perspective is to consider the sets of equations (\ref{discrete1}) as one single system of equations driven by the matrix 
\begin{eqnarray}
T=\begin{pmatrix}
T_+  & T_-\\
T_- & T_+
\end{pmatrix}.
\end{eqnarray}
The latter is simply the (unsigned) transition matrix of a larger graph, composed of $2N$ nodes (each node appears twice, once for positive walkers, and once for negative walkers), with adjacency matrix
\begin{eqnarray}
\label{largerA}
A=\begin{pmatrix}
A_+  & A_-\\
A_- & A_+
\end{pmatrix}.
\end{eqnarray}
The whole diffusive process can thus be understood in terms of the spectral properties of that transition matrix. The system of equations decouples when performing the following change of variable: for each $i$, we define $\Delta_i = n_{+;i} - n_{-;i}$ and $n_i = n_{+;i} + n_{-;i}$, where the determinant of the transformation matrix is $2$, leading to the transition matrix 
 \begin{eqnarray}
T^{'}=\begin{pmatrix}
T_s  & 0\\
0 & T_u
\end{pmatrix},
\end{eqnarray}
for these variables. The spectral properties of $T$ are therefore determined by those of the two matrices $T_s$ and $T_u$, with well-known special cases, e.g., when the graph is disconnected or (strongly) structurally balanced \cite{tian2024spreading}. 

\subsection{Walk lengths}

The strong walk can be equipped with a teleportation mechanism in order to keep walkers concentrated in the neighborhood of a target node. 
Given a teleportation probability $(1-p_s)$, we consider the expected walk length before a teleportation. We know that if $p_s = 1$, the strong walk will have infinity as the expected length, thus we assume $p_s < 1$ here. 
For each node $i$, if the walk has length $k$, it means that teleportation does not happen at the first $(k-1)$ time steps but the $k$th time step, thus the number of steps that teleportation does not occur follows a negative binomial distribution with success probability being $1-p_s$ and the number of success being $1$. Hence, the expected walk length is the sum of the expected steps that teleportation does not happen plus $1$, i.e., 
\begin{align*}
    \mathbb{E}[l|X_0=i] = \frac{p_s}{1-p_s} + 1 = \frac{1}{1-p_s}. 
\end{align*}

\section{Signed network similarity matrix}

In the case of unsigned networks, most kernels are based on the principle that similar nodes tend to be connected by several, ideally short walks, which can be captured by the flow of walkers between them. In the case of signed graphs, one needs to distinguish between the type of walks, as some of them  will make a pair of nodes more similar, while others will make that pair more dissimilar. Hence, a modelling choice has to be made to define ``positive'' versus ``negative'' walks. 

Let us start with an intuition based on the notion of strong balance. A graph is said to be strongly balanced if it does not possess any cycle with an odd number of negative edges. Equivalently, this graph can be partitioned into two sets, say $A$ and $B$, of nodes such that the positive edges are within each cluster, and the negative edges are between the two. This also implies that any walk between two nodes of the same set, say $A$, possesses an even number of negative edges, while any walk between two nodes of different sets possesses an odd number of negative edges. 

Given this extra ingredient, we follow the steps leading us to the similarity matrix inspired by random walks with restart. For each node $i$, we consider a process where with probability $p_s$, walkers may either jump to one of its neighbouring nodes, and possibly change sign, or teleport back to the pinned node with probability $1-p_s$. During the teleportation, the walker automatically becomes positive, independently of its sign. This choice is motivated by the fact that a node should be similar to itself. As before, the process defines for each node $i$ a vector, but it is now a $2N$ vector, where the first $N$ entries encode the density of positive walkers (arriving through walks with an even number of negative edges, or positive walks), while the last $N$ entries encode the density of negative walkers (arriving through negative walks). 

The walker begins as a positive walker on some node $i$, $\textbf{e}_{i}$, a $2N$ vector.  
The state of the walker after $k$ steps, ${\bf u}_{k}$ can be written in terms of the walker after $k-1$ steps ${\bf u}_{k-1}$ as, 
\begin{eqnarray}
{\bf u}_{k} = (1-p_s) {\bf e}_i + p_{s}{\bf u}_{k-1} T,
\end{eqnarray}
according to the dynamics defined by the transition matrix $T$. The stationary state of this process is defined by
 \begin{eqnarray}
{\bf n} = (1-p_s) {\bf e}_i (I - p_s T)^{-1}. 
\label{equ:kernel-strong}
\end{eqnarray} 
The first $N$ entries of {\bf n}  encode the flow of walkers on positive walks from $i$ to the other nodes, hence capturing similarity between the nodes. The last $N$ entries encode the flow of walkers on negative walks from $i$ to the other nodes, hence encoding dissimilarity between the nodes. 
More precisely, stacking 
the first $N$ elements of this vector
defines the $i$-th row of a similarity matrix $\kappa_+$.
Stacking 
the last $N$ elements of this vector 
defines the $i$-th row a dissimilarity matrix $\kappa_-$. 
The two matrices can be seen as the positive and negative adjacency matrices of a graph where weights involve longer walks that explore larger area of the network when $p_{s}$ increases.
Taking the differences between the matrices defines a signed similarity matrix, which compares the relative importance of positive and negative walks  between nodes. Note that in Eq.(\ref{equ:kernel-strong}), we give the same importance to each node, in contrast to the Eq.(\ref{kernel1}), as there is no straightforward way to make the similarity matrix symmetric by a weighting of the nodes. Other choices could be done, for instance, by weighting the nodes by their degree. In practice, for the experiments where we need to find the entire kernel, we symmetrize the signed similarity matrix after calculating it by adding the resulting matrix to its transpose.

\section{Weak walks}

\subsection{Definition and basic properties}
\begin{figure}[]
\centering
\includegraphics[width=0.8\linewidth]{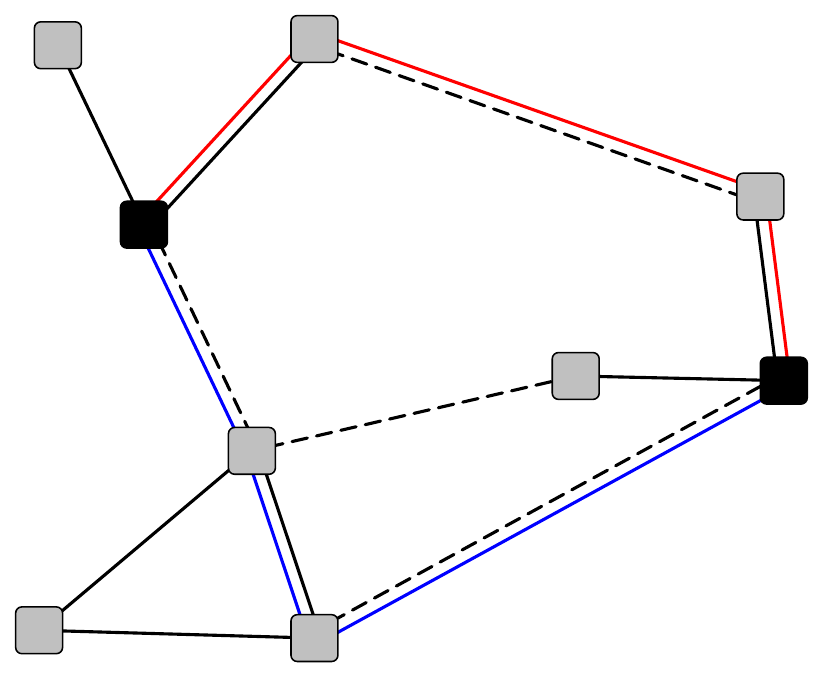}
\caption{When estimating the similarity between two nodes, here colored in black, strong walks may lead to conflictual signals when the network is composed of more than 2 clusters, here illustrated via a positive walk in blue and a negative walk in red (solid: positive; dashed: negative). Weak walks only consider walks with at most one negative edge and thus properly consider the two nodes as dissimilar. }
\label{fig:signed}       
\end{figure}

The walk (\ref{discrete1}) appears to be appropriate to uncover the similarity of two nodes from a strong balance perspective, as walks with an even number of negative edges will increase the similarity between the nodes, while walks with an odd number of negative edges will decrease their similarity. However, this is not the case when searching similar nodes based on the notion of weak balance; see Figure \ref{fig:signed}. A graph is said to be weakly balanced if it does not possess cycles with one single negative edge. Weakly balanced
graphs are k-clusterable, meaning that they can be partitioned into k factions, where the negative edges are between the factions and the positive edges are within the factions. 
The condition via the appearance of one single negative edge suggests to define the dissimilarity between two nodes by the existence of walks with one single negative edge between them. Any walks involving more than one negative edge are instead considered uninformative and should not be considered when estimating the similarity between two nodes. 

These arguments suggest to modify the strong walk defined by imposing that a walker can become negative only once and that it disappears when taking a negative edge a second time. This argument causes an asymmetry in the node states, as positive and negative walkers now play a different role. The dynamics is  encoded by the enlarged transition matrix 
\begin{eqnarray}
T_{w} =	\begin{pmatrix}
T_+ &T_-\\
0	&T_+
\end{pmatrix},
\label{weakwalkmat}
\end{eqnarray}
It is easy to show that it will keep the same structure, after several iterations, and that:
\begin{eqnarray}
T_{w}^n=\begin{pmatrix}
T^n_+  & U_n\\
0 & T^n_+
\end{pmatrix},
\end{eqnarray}
where $U_n = \sum_{i = 0}^{n-1}\binom{n-1}{i}(T_{+})^{i}(T_{-})(T_{+})^{n-1-i}$. As defined here, the process leads to a leak of probability, as walkers disappear after taking a negative edge for the second time, leading to a vanishing number of walkers asymptotically (except in trivial cases). 

A natural way to deal with this leakage is to adopt a random walk with restart. We assume that a negative walker teleports back to a given target node when taking a negative edge. 
In contrast with the case of the strong walk, an additional teleportation is not necessary for the walks to remain finite before the restart, unless in trivial situations (e.g., a connected component without a single negative edge). Moreover, the length of the walks before the restart is not controlled by a parameter, but controlled by the structure of the signed network.

Suppose we start with $1$ positive walker on node x, given by the initial state vector, $\textbf{e}_{x}$, where this is a vector of length $2N$. Then, the random walk with restart would be defined by the matrix: 

\begin{eqnarray}
T_{x} =	\begin{pmatrix}
T_+ & T_-\\
M_{x} & T_+
\end{pmatrix},
\label{tx}
\end{eqnarray}
where $M_{x}$ is equivalently the zero matrix except column $x$, which is equivalent to the vector $\textbf{v}$ where: 

\begin{equation}
v_{i} = \sum_{j}T_{-;ij},
\end{equation}
such that matrix $T$ is now stochastic and has row sum equal to $1$. The negative walker will return to node $x$ as a positive walker when it encounters another negative edge. The evolution can be determined by taking the product of the initial state vector $\textbf{e}_{x}$ and the matrix $T_{x}$. 

The walk can also be modified to have an additional teleportation parameter, $p_w$, where each walker will teleport back to the pinned node with additional probability $1-p_w$ as a positive walker, in addition to the restart in the case of a negative walker encountering a negative edge.
This results in an increase of the teleportation rate as compared to the strong walk case, thus a decrease in the expected walk length, as we will show. In this framework, we can then construct a signed similarity matrix in the very same way as for strong walks.   For a random walk starting as a positive one on node $x$, the state of the weak walker after $k$ steps, ${\bf u}_{k}$ can be written in terms of the walker after $k-1$ steps ${\bf u}_{k-1}$ as, 
\begin{eqnarray}
{\bf u}_{k} = (1-p_w) {\bf e}_x + p_{w}{\bf u}_{k-1} T_{x}, 
\label{equ:state-weak}
\end{eqnarray}
in much the same fashion as in Equation \eqref{equ:kernel-strong}. 

\subsection{Walk length}
We now characterise the weak walks from the perspective of expected walk length. 
We consider a teleportation probability $1-p_w$. 
In this case, teleportation is not the only reason for the random walker to stop (or restart), where traversing negative edges for the second time can also trigger it. Hence, we cannot have a network-independent distribution, but need to consider the network structure. 
Particularly, we consider the enlarged transition matrix \eqref{weakwalkmat}.  
We encode the expected number of visits in the matrix $Q\in\mathbb{R}^{2N\times 2N}$, where $N$ dummy nodes are created to separate the cases when the the node is visited via paths of zero or one negative edge. Then, 
\begin{align*}
    Q = \sum_{k=0}^{\infty}p_w^kT_w^k = (I - p_wT_w)^{-1},
\end{align*}
where $I$ is the identity matrix. 


Hence, the expected walk length starting from node $i$ is
\begin{align*}
    \mathbb{E}[l|X_0 = i]=\sum_{j=1}^{2N}Q_{ij}.
\end{align*}
Note that we only count the first $N$ rows in $Q$. 
Therefore, the expected walk length for weak walk is 
\begin{align*}
    \mathbb{E}[l] = \sum_{i=1}^{N}\mathbb{E}[l|X_0 = i]P(X_0=i).
\end{align*}

\begin{figure*}
    \begin{tabular}{ccc}
        \includegraphics[width=.3\linewidth]{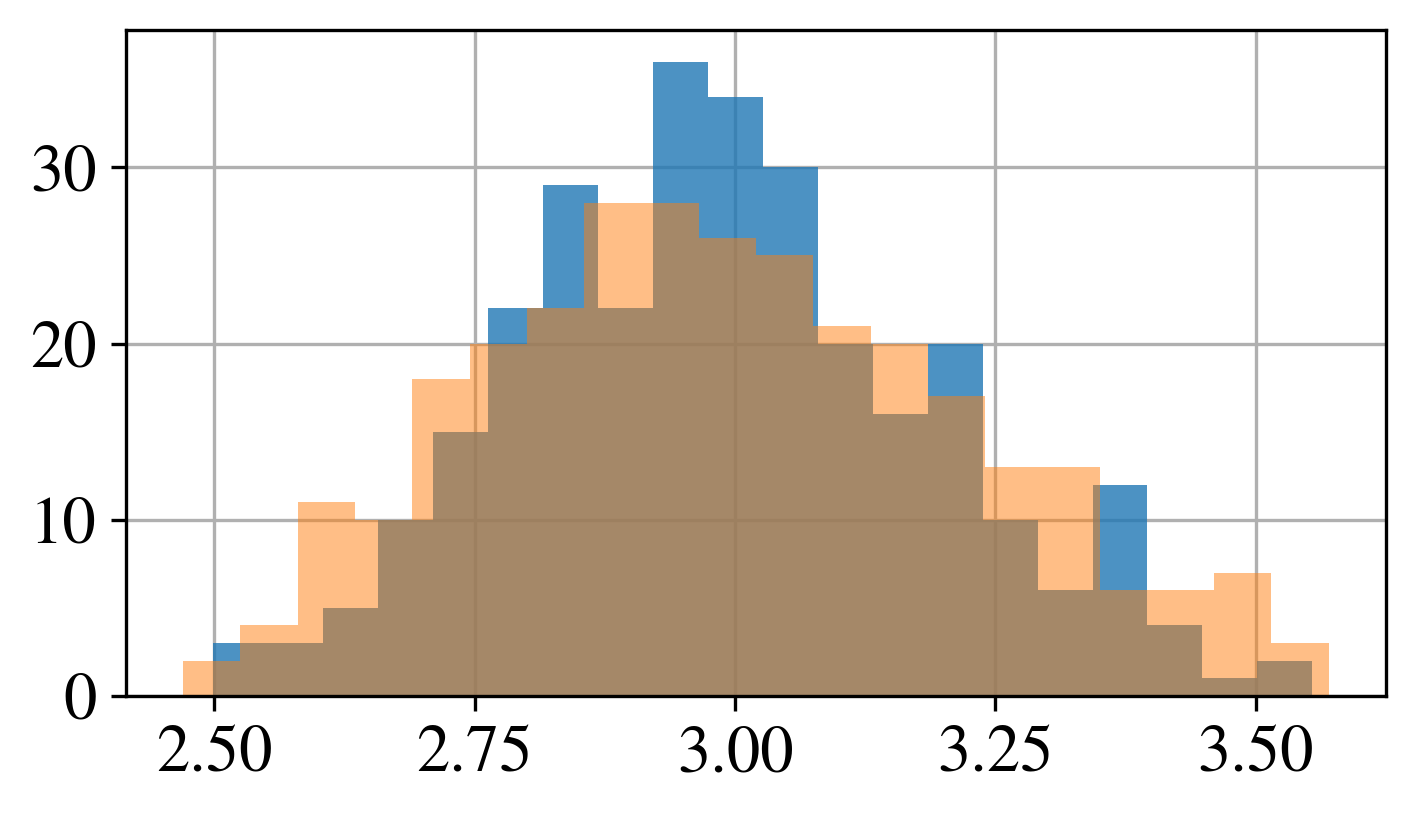} & \includegraphics[width=.3\linewidth]{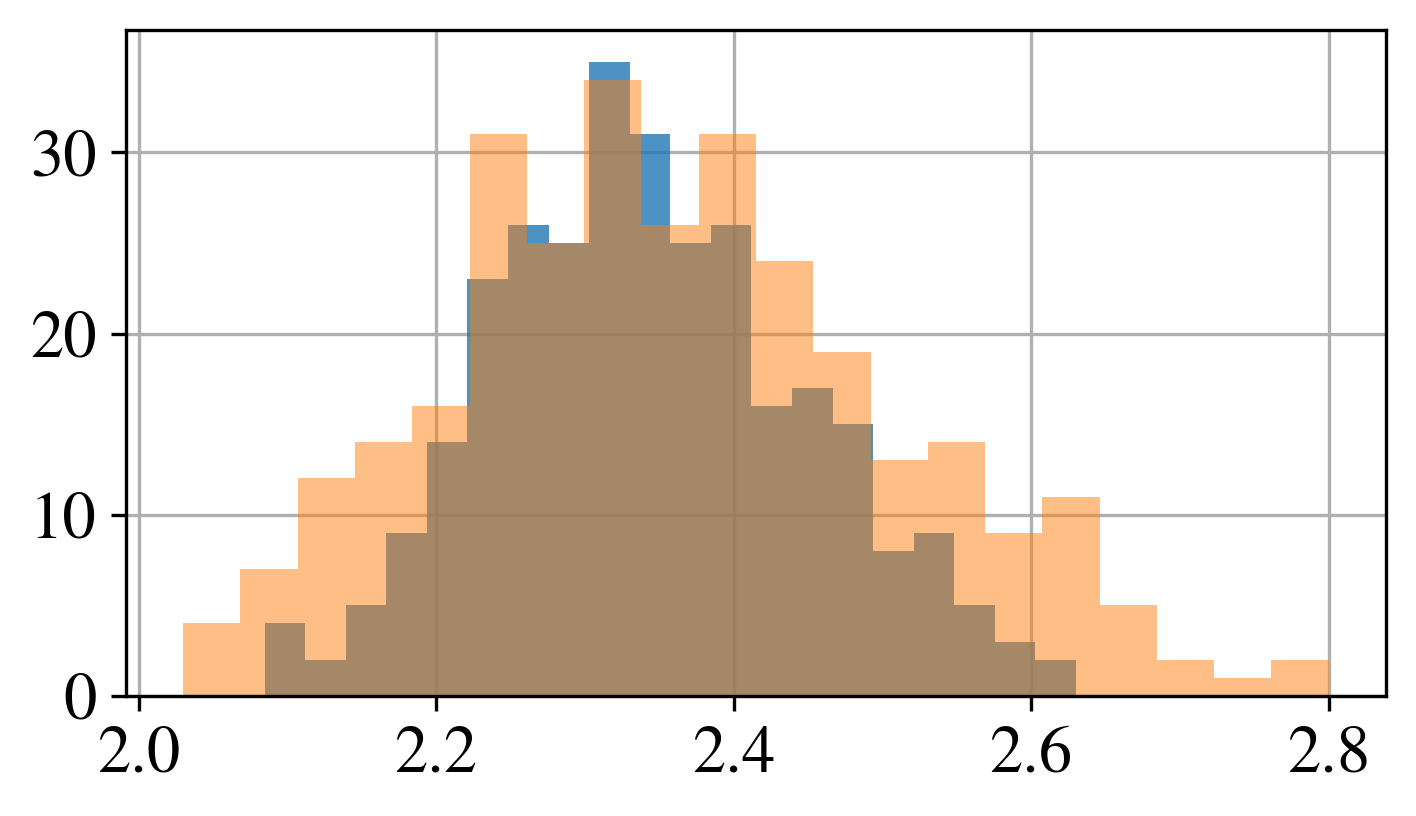} & \includegraphics[width=.3\linewidth]{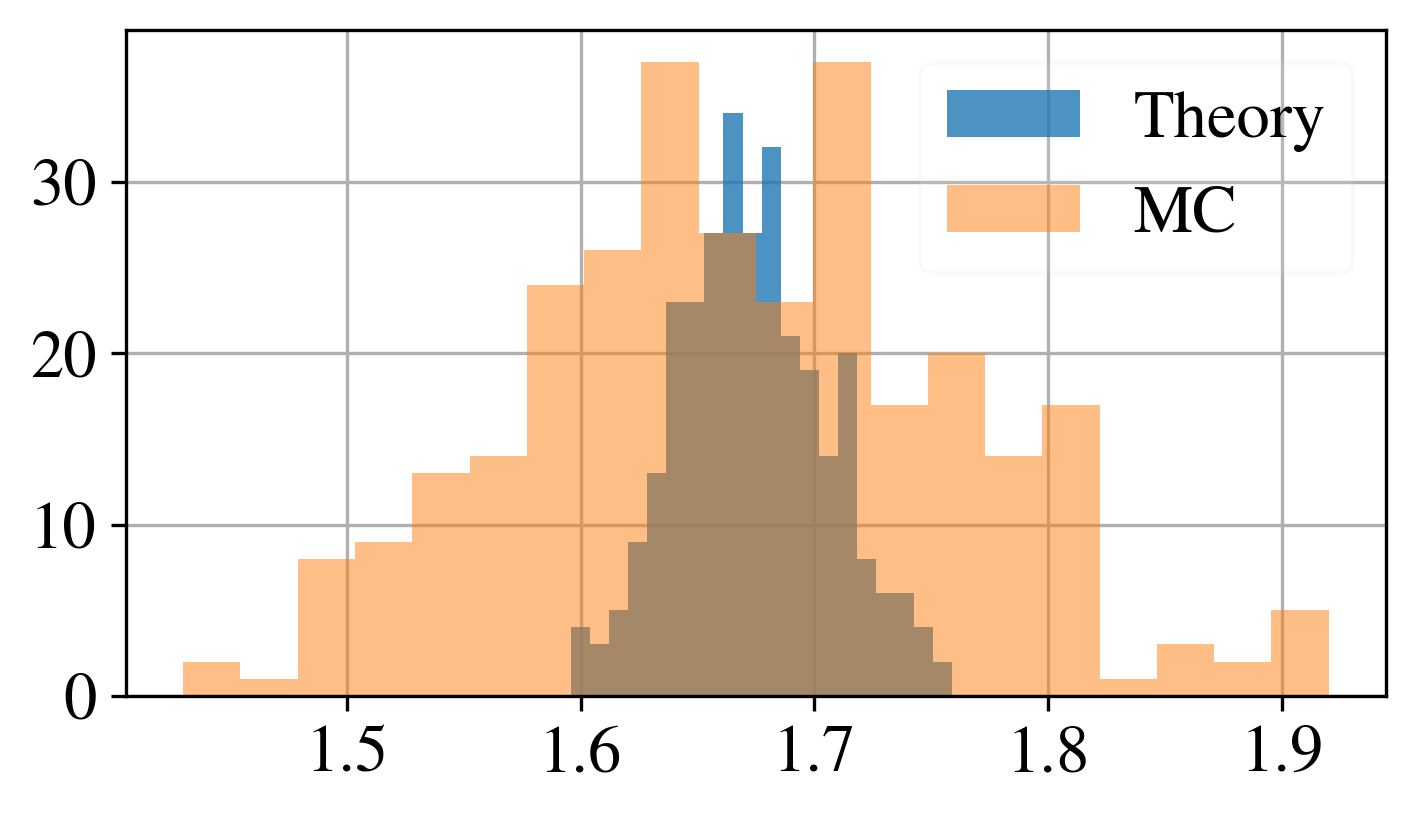}
    \end{tabular}
    \caption{Distribution of the expected walk lengths starting from different nodes, both from the theoretical results in blue and Monte-Carlo simulations in orange on $3$ community symmetric SSBM with $100$ nodes in each community, $p=0.05$ and different levels of teleportation (left: $p_w=1$; middle: $p_w=0.8$; right: $p_w=0.5$, where the strong walk with $p_s=0.8, 0.5$ has expected walk length $5, 2$, respectively).}
    \label{fig:ww-3com-ps}
\end{figure*}

We verify the theoretical results with Monte Carlo simulations on Signed Stochastic Block Models (SSBMs, see section \ref{sec:SSBM} for details), where edges occur with probability $0.05$ and do not flip sign from the initial configuration; see Figure \ref{fig:ww-3com-ps}. We consider a uniform initial distribution, repeat the random walks starting from each node for $n_s=100$ times and obtain the expected walk length up to either teleportation happens or it traverses the second negative edge.

\section{Applications}
In this section we present an application of the weak walk method to signed network clustering. We first show the difference between the resulting signed similarity matrices for the weak and strong walks on two small signed network examples, for which ground truth clusters exist. Next, we compare the weak walk against two other methods by performing semi-supervised clustering on a series of synthetic SSBM networks, with varying levels of noise, community numbers, and link densities.  Specifically, we compare the weak walk against (1) the strong walk and (2) the Signed Random Walk with Restart (SRWR), a three-parameter signed network walk designed for semi-supervised clustering which generates a ranking of nodes in signed networks~\cite{7837935}. The SRWR model considers a random walker that changes its sign as it takes positive or negative edges according to two parameters, as well as a teleportation parameter, and exhibits high accuracy at signed link prediction.  The semi-supervised clustering for each random walk is performed in much the same fashion as with PageRank~\cite{langville2006google}; we begin with (positive) walkers on a set of labeled seed nodes. We use the similarity vector generated from the weak, strong and SRWR walks to obtain a ranking of each of the other nodes in the graph. Each unclassified node is assigned a label depending on in which walk it ranked the highest, inheriting the label of that seed node.  We use the parameter-free version of the weak walk with no teleportation ($p_w = 1$).  The strong walk has one one parameter to adjust, the teleportation parameter ($p_s$), while SRWR has three parameters: two parameters which control if the walker changes sign when it takes a negative edge or not, as well as a teleportation parameter. In these experiments, we sweep through the parameter space for the strong walk and the SRWR and take the parameters with the best results. We show that the weak walk outperforms the other two methods in certain cases, despite being parameter-free. To evaluate the accuracy of the clustering assignment obtained in each case we use the Adjusted Rand Index (ARI)~\cite{gates2017impact} to compare against the ground truth clusters.

\subsection{Computational details}
In the experiments that follow, we make the choice to construct the signed similarity matrices for both the strong and weak walks by taking the state of the walk vector at $k = n$, where $n$ is large enough that the walk is approaching a stationary state. This can be computationally written as an iterative process allowing us to solve for the similarity matrix as a whole in one computation, which is more efficient than inverting a matrix as required by Equation \eqref{equ:kernel-strong}. As well, the matrix given in Equation \eqref{equ:state-weak} is specific to the starting node $x$, requiring $N$ matrices to be inverted. See section \ref{sec:algorithm} for a description of the algorithm. 
Specifically, if the state of the walker (strong or weak) is given by ${\bf u}_{k}$, the similarity between node $x$ and any other node is given by the vector $s_{x}$: 
\begin{equation}
    s_{x} = u_{k =n}[0:N] - u_{k =n}[N+1:2N],
\label{similarity_eq}
\end{equation}
since we subtract the negative walks from the positive. For the experiments on the empirical networks for which we calculate the entire similarity matrices, the kernel is made symmetric by adding it to its transpose. 

\subsection{Highland Tribes}
\begin{figure*}[]
    \centering
    \begin{tabular}{cc}
        \includegraphics[width=.3\linewidth]{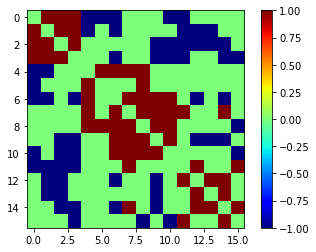}
        \includegraphics[width=.3\linewidth]{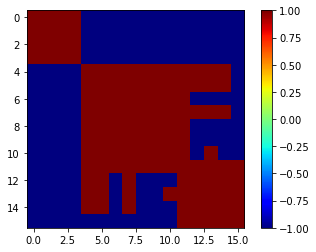} & \includegraphics[width=.3\linewidth]{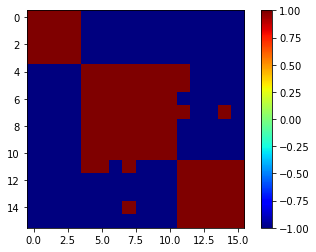}
    \end{tabular}
    \caption{(left) Adjacency matrix for the Highland Tribes network. (centre) Strong walk kernel resulting from a 500 step walk on the Highland Tribes network with teleportation parameter $p_{s} = 0.7$. (right) Weak walk kernel resulting from a 500 step walk with no teleportation $p_{w} = 1.0$. Positive and negative values have been set to +1 or -1 respectively for clearer visualization. }
    \label{fig:tribe_kernels}
\end{figure*}

There are very few empirical signed networks available that are labeled with ground truth cluster assignments. We show here two small examples with labeled data to demonstrate the distinction between the strong and walk walk kernels on these networks. The first network comes from \cite{read1954cultures}, and represents the antagonistic or friendly relationships between 16 tribes of the Eastern Central Highlands of New Guinea, represented by edge weights of -1 or +1 respectively. The network can be clustered into three communities, which characterise the higher order groupings of the society~\cite{hage1983structural}. Figure \ref{fig:tribe_kernels} gives the adjacency matrix for the Highland Tribes network, where the nodes have been ordered by their community groupings. This network is not perfectly balanced, and as it has more than two communities, we expect that the weak walk will outperform the strong walk at locating the three communities.  The adjacency matrix can be combined with other methods such as the  spin-glass community detection method for signed networks based on simulated annealing introduced in ~\cite{traag2009community} to obtain the community assignments in an unsupervised fashion, which can be compared against the ground truth community assignments using the ARI score. The resulting strong and weak walk kernels generated by this network with 500 steps are also given in Figure \ref{fig:tribe_kernels}. The weak walk is parameter free, with no teleportation. For the strong walk, we choose the teleportation parameter by inputting the kernel into the spin-glass algorithm, and comparing the community assignments against the ground truth. We take the teleportation parameter that yields the best ARI score, corresponding to $p_s= 0.7$.  The strong walk kernel exhibits far more incorrect signs as compared to the weak walk: in particular it exhibits several positive connections between the second and third communities. 
Like the adjacency matrix, these kernels can be combined with the spin-glass algorithm to obtain unsupervised community assignments. 
In this case, the graph is quite small, dense and almost balanced, and so the adjacency matrix, weak walk kernel and strong walk kernel all produce an ARI of 1.0 when used as input into the spin-glass algorithm.

\subsection{Sampson's Monastery}
\begin{figure*}[]
    \centering
    \begin{tabular}{cc}
        \includegraphics[width=.3\linewidth]{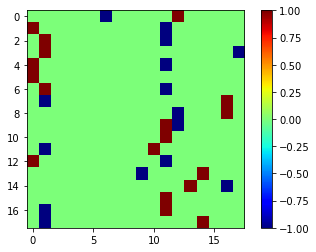}
        \includegraphics[width=.3\linewidth]{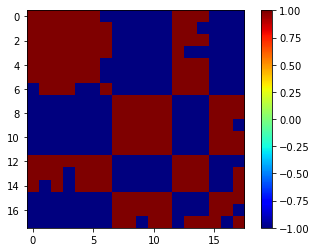} & \includegraphics[width=.3\linewidth]{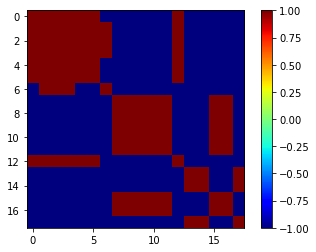}
    \end{tabular}
    \caption{(left) Adjacency matrix for the Sampson's Monastary network. (centre) Strong walk kernel resulting from a 500 step walk on the Highland Tribes network with teleportation parameter $p_s = 0.8$. (right) Weak walk kernel resulting from a 500 step walk with no teleportation. Positive and negative values have been set to +1 or -1 respectively for clearer visualization.}
    \label{fig:sampson_kernels}
\end{figure*}

In this section we present another example of a small empirical signed network with ground truth labels. The Sampson's Monastery network comes from an ethnographic study of the community structure within a New England monastery, where each relationship between the 18 novice monks has been designated as positive (friendly) or negative (antagonistic). Sampson, the ethnographer, divided the men into four groups according to the community structure he observed~\cite{sampson1968novitiate}. Although this example also represents a small network, the graph is more sparse than the Highland Tribes example, and has more noise in terms of edge signs that violate structural balance. Figure \ref{fig:sampson_kernels} gives the network adjacency matrix, with the nodes arranged according to the four ground truth communities. Again we expect that since this network has more than two communities, the weak walk will outperform the strong walk at identifying the four communities. The resulting strong and weak walk kernels generated by this network with 500 steps are also given in Figure \ref{fig:sampson_kernels}. Again, we use the parameter-free weak walk with no teleportation, while for the strong walk, we choose the teleportation parameter of $p_s = 0.8$, which achieves the highest ARI score when combined with the spin-glass algorithm.  The strong walk kernel exhibits far more incorrect signs as compared to the weak walk, generating several positive connections between the second and third communities, and the third and fourth communities. When combined with the community detection spin-glass algorithm, the parameter-free weak walk kernel obtains the best ARI (0.639) compared to the the adjacency matrix (0.442) and the strong walk kernel (0.518).

\section{Application: semi-supervised clustering}
In this section, we use the weak walk kernel to perform semi-supervised clustering of several different synthetic SSBMs with community structure. We compare against the kernel generated by the strong walk, and the SRWR. In these experiments we take the parameter-free version of the weak walk, with no teleportation, while we sweep through the parameter space for the strong walk and the SRWR and take the parameters which achieve the highest ARI. Consequently, these methods perform better than the weak walk in some cases, but it may be less useful in practice, since the ideal parameters are not known a priori and indeed tend to change with each iteration of the SSBM, even when the SSBM parameters are held constant. 

\subsection{Strong Balance: Two-Community SSBMs}
\begin{figure}[h!]
\centering
\includegraphics[width=0.8\linewidth]{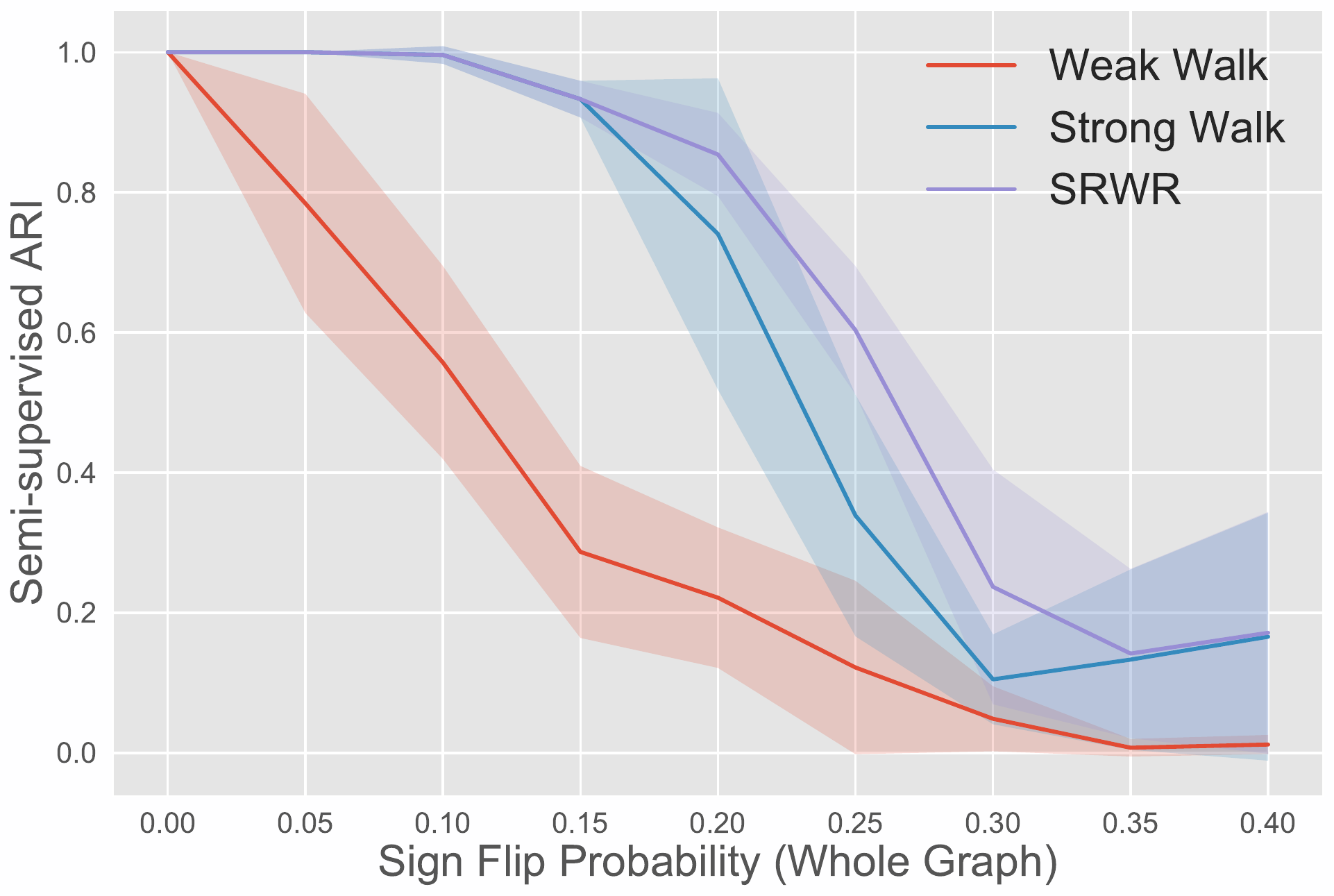}
\caption{ARI from the semi-supervised clustering for the weak walk, strong walk and SRWR on 2 community SSBMs (strong balance).}
\label{fig:2com}       
\end{figure}

Here, we look at 2 community SSBMs with 100 nodes in each community, and 0.05 probability of link existence between each pair of nodes. We perform the semi-supervised clustering with varying levels of noise on the signs of the links, via the sign flip parameter $\nu$ (see section \ref{sec:SSBM} for a description of the noise parameter). For the strong and weak walks, we take the state vector at iteration $k = 100$, and set for SRWR the max iterations = 100. The other parameters for the SRWR and strong walk are chosen by parameters sweep and change for each iteration of the graph. We take the mean and standard deviation over 10 graph realizations, and the results are given in Figure \ref{fig:2com}. We observe that the SRWR performs the best, a feature to be balanced with the fact that it has the most parameters. The strong walk exhibits improvement over the weak walk; this is expected, as the weak walk is designed to outperform the strong walk in the case when there are more than two communities, and the strong walk has the advantage of teleportation. 

\subsection{Weak Balance: Three-Community SSBMs}

\begin{figure*}[]
    \centering
    \begin{tabular}{cc}
        \includegraphics[width=.33\linewidth]{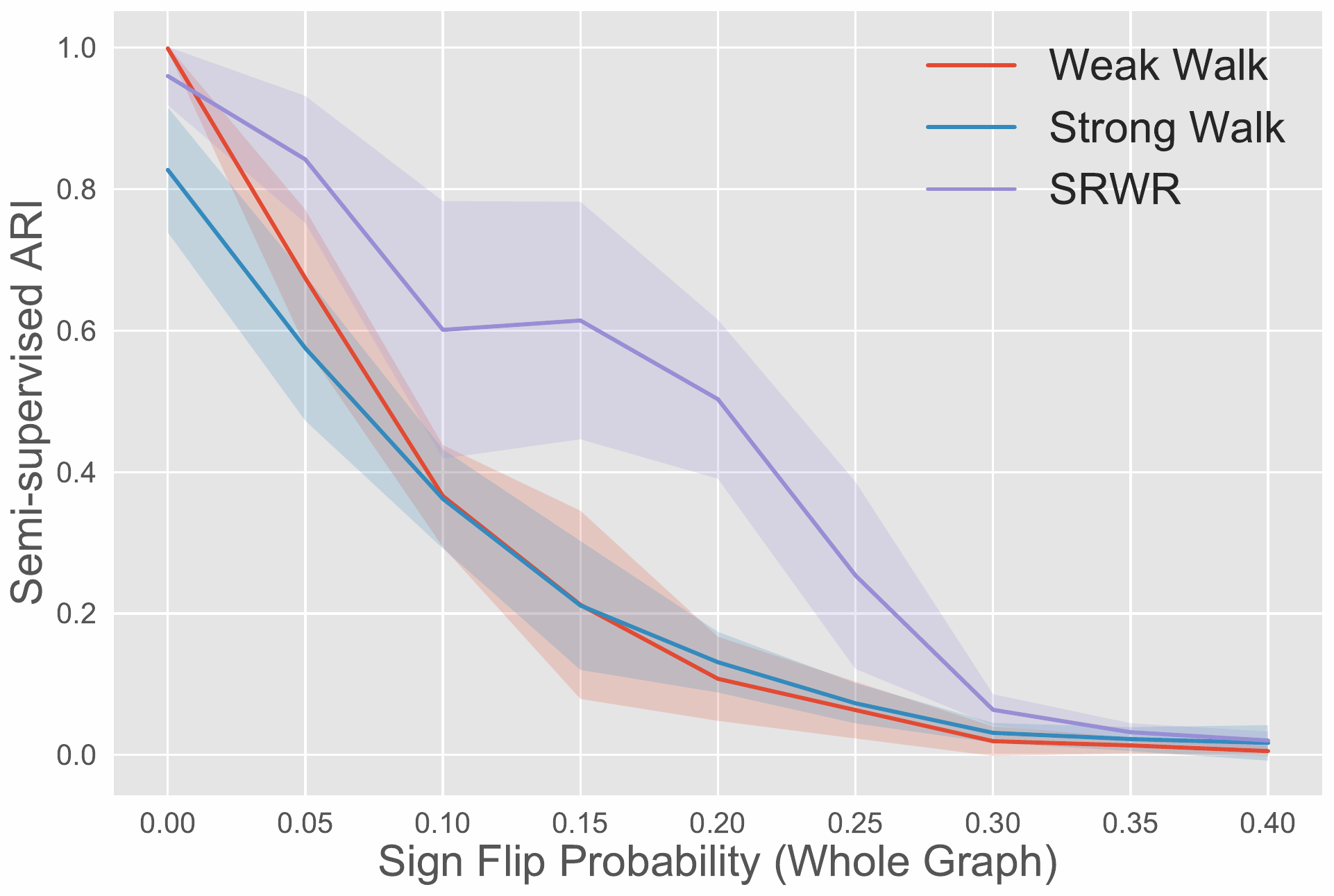}
        \includegraphics[width=.33\linewidth]{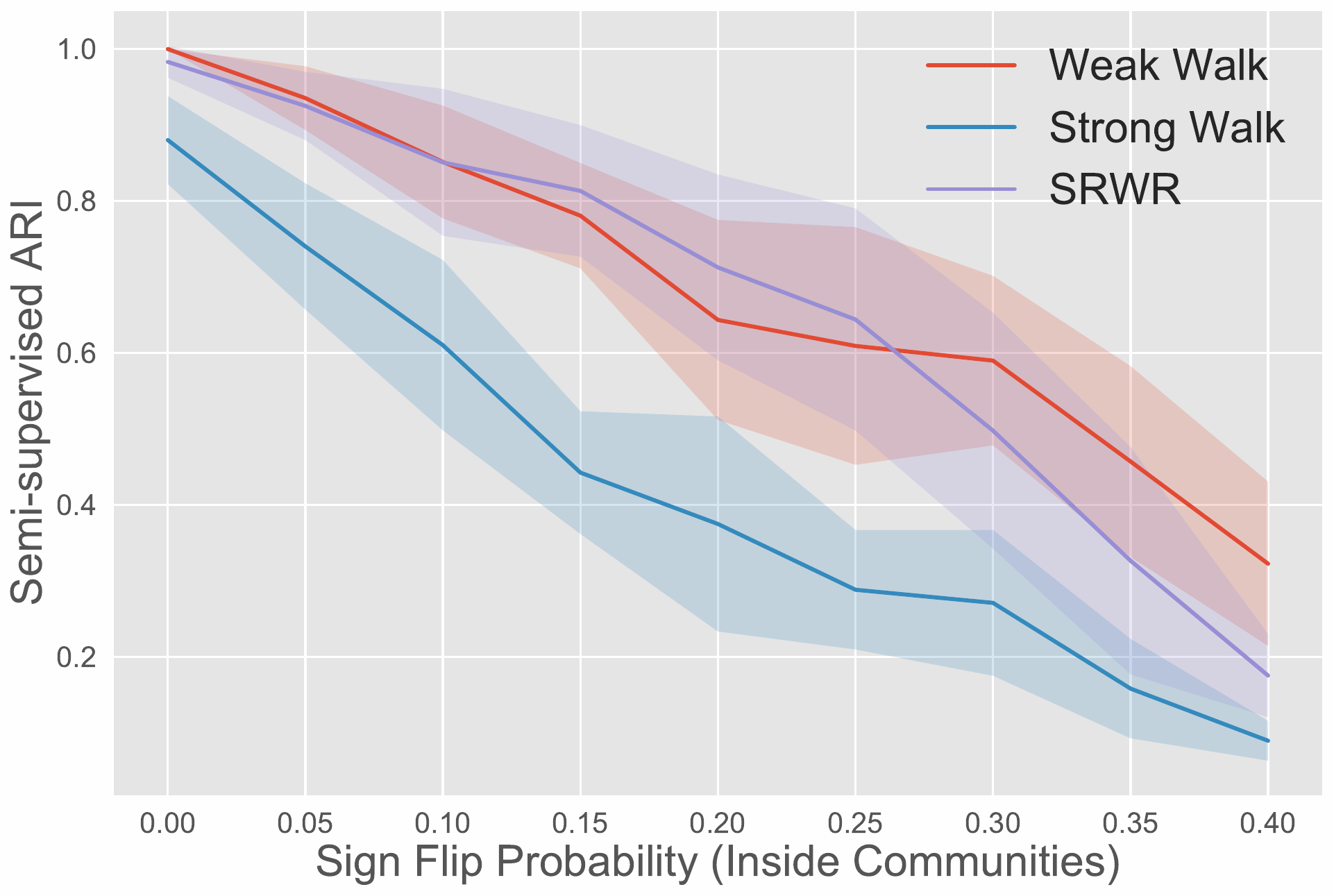} & \includegraphics[width=.33\linewidth]{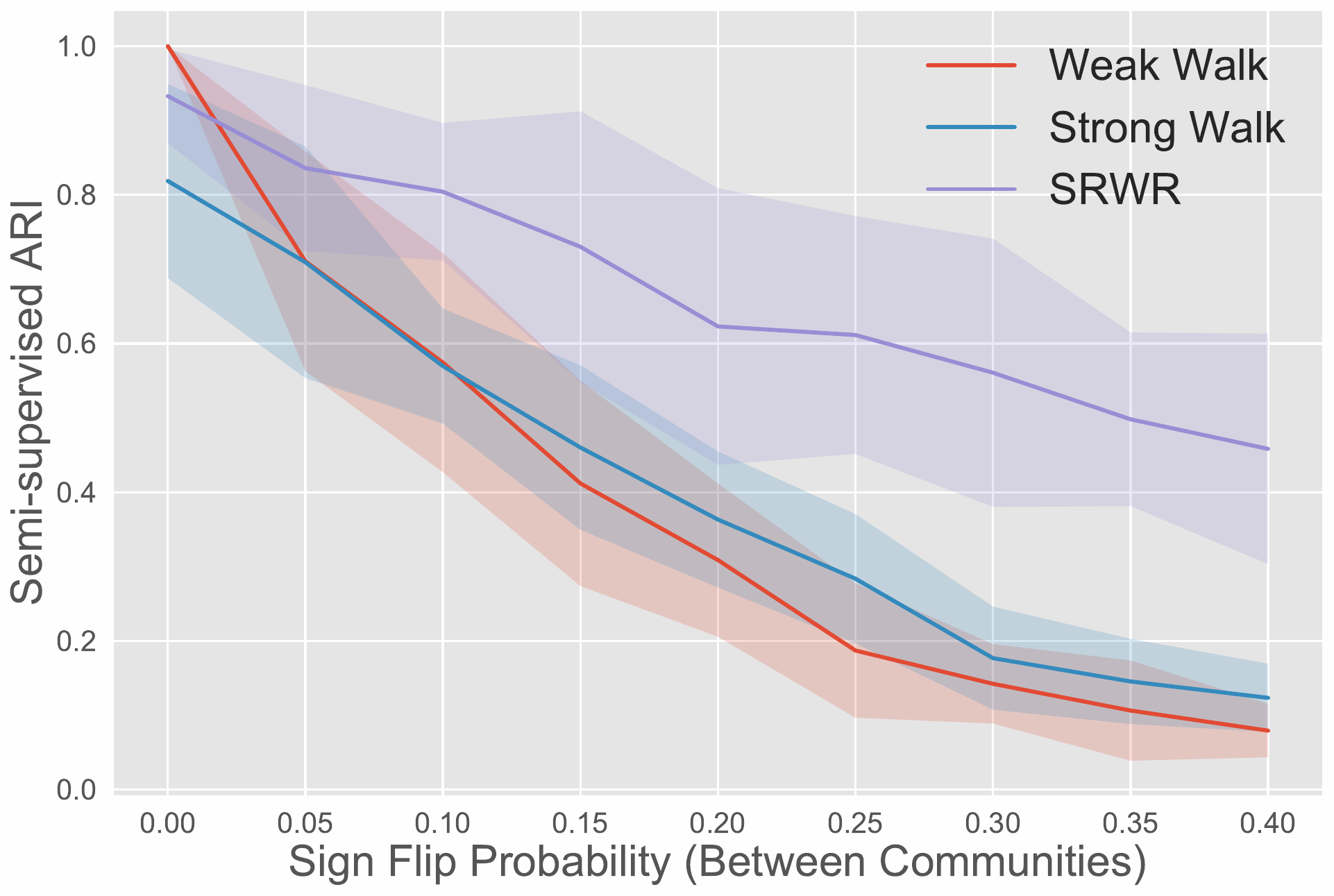}
    \end{tabular}
    \caption{ARI from the semi-supervised clustering for the weak walk, strong walk and SRWR on 3 community SSBMs (weak balance). (left) Noise level general to whole graph. (middle) Noise is only inside communities. (right). Noise is only between communities. }
    \label{fig:3com}
\end{figure*}

Here, we look at 3 community SSBMs with 100 nodes in each community, and 0.05 probability of link existence between each pair of nodes. We perform the semi-supervised clustering with varying levels of noise on the signs of the links, as in the previous example. For the strong and weak walks, we take the state vector at iteration $k = 100$, and set for SRWR the max iterations = 100. As before, the other parameters are chosen by parameters sweep and change for each iteration of the graph. We take the mean and standard deviation over 10 graph realizations, the results are given in Figure \ref{fig:3com} (left). We observe that the SRWR again outperforms the other two methods. The strong walk and weak walk exhibit similar performance, despite the strong walk having one extra parameter. In different empirical networks, the noise may occur along positive or negative edges, with different probabilities. Consequently, we investigate the results when the noise on the signs of the edges is restricted to either inside the communities by varying $\nu_{inside}$ and setting $\nu_{between} = 0$, or between communities by varying $\nu_{between}$ and setting $\nu_{inside} = 0$ (see section \ref{sec:SSBM} for details). The weak walk is much more robust to noise inside communities (Figure \ref{fig:3com} (centre)), and indeed performs almost as well as the SRWR in this case. When the noise is between the communities (Figure \ref{fig:3com} (right)), the weak walk is less robust and the SRWR outperforms both the strong and weak walks.

\subsection{Weak Balance: Six-Community SSBMs}

\begin{figure*}[]
    \centering
    \begin{tabular}{cc}
        \includegraphics[width=.33\linewidth]{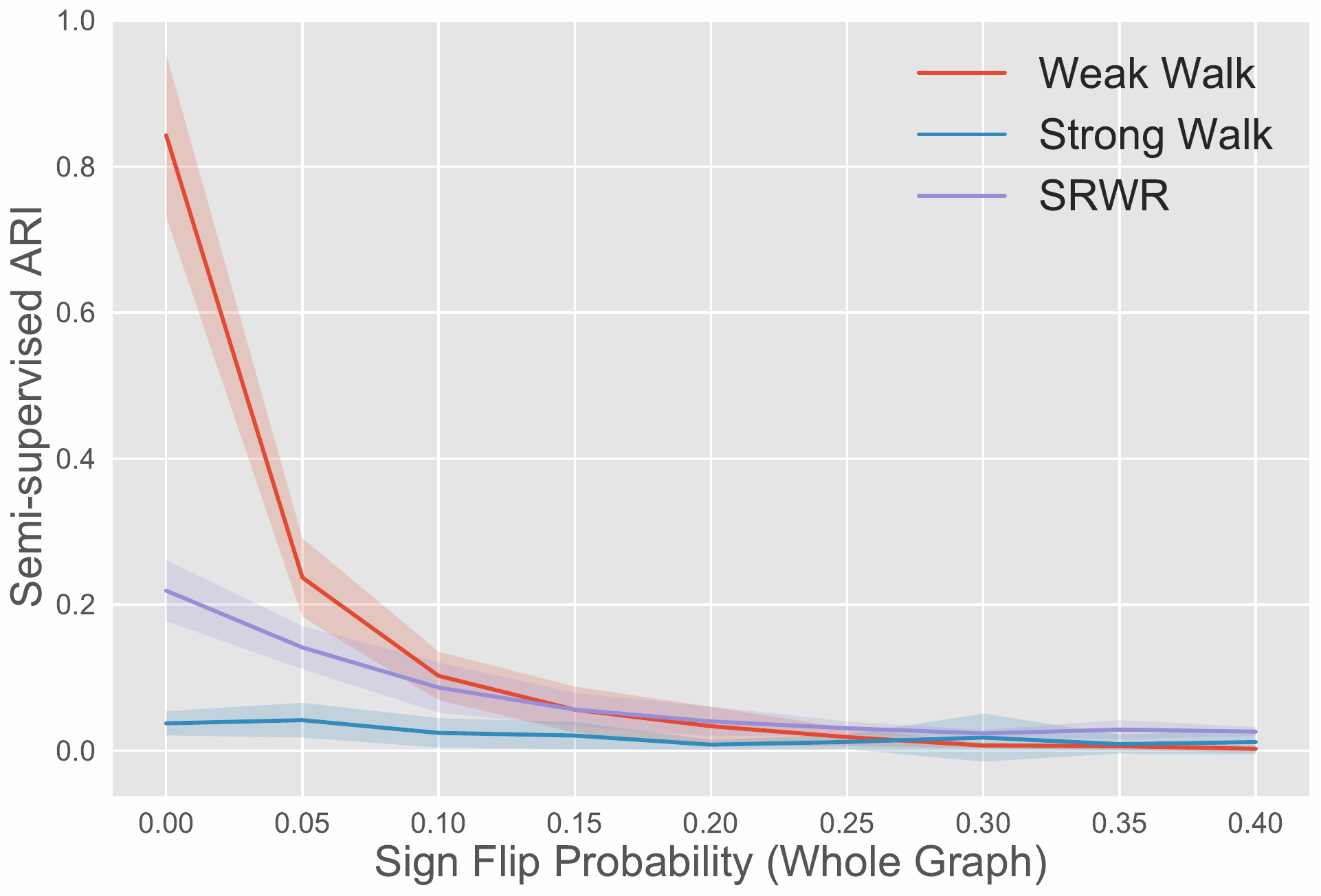}
        \includegraphics[width=.33\linewidth]{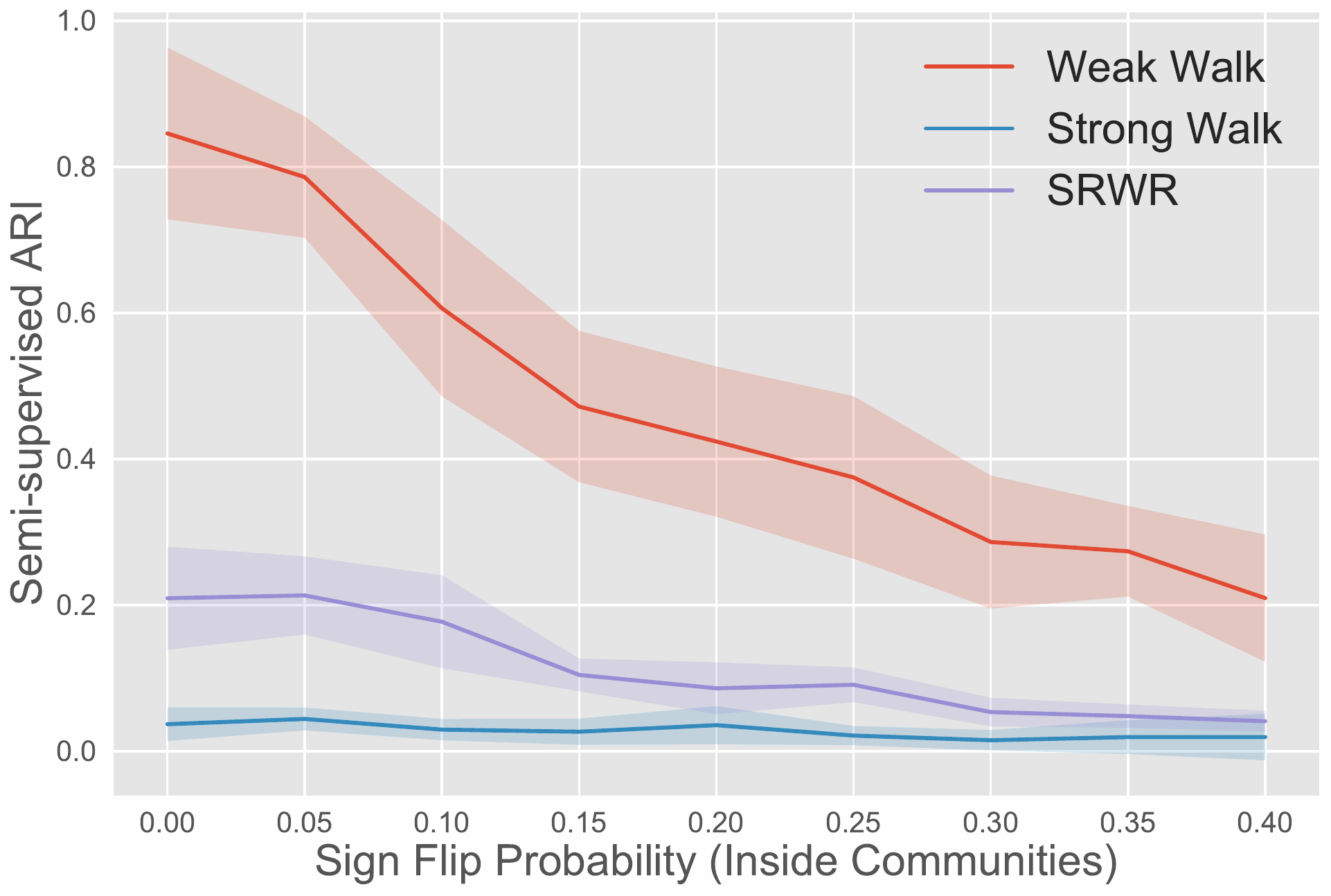} & \includegraphics[width=.33\linewidth]{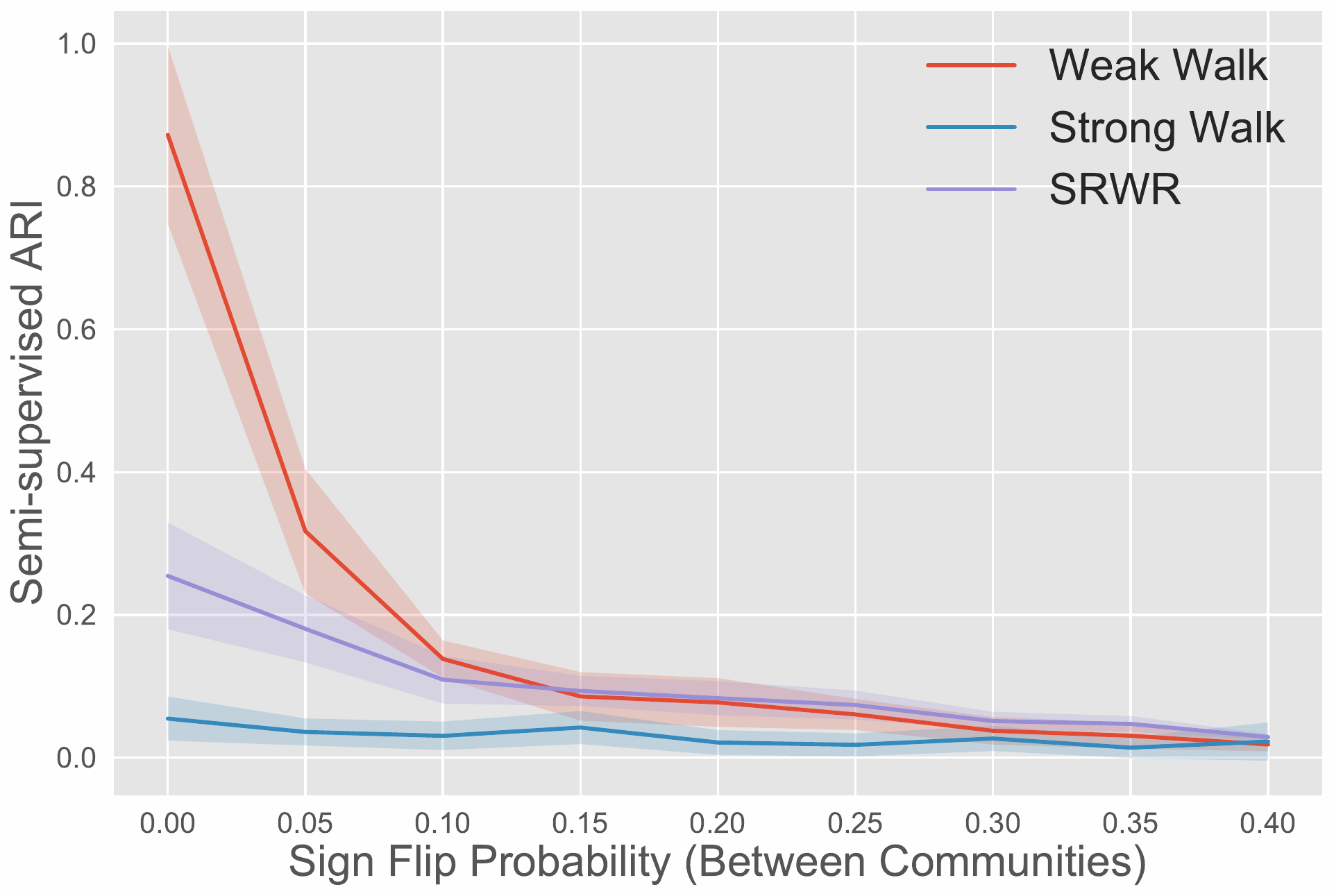}
    \end{tabular}
    \caption{ARI from the semi-supervised clustering for the weak walk, strong walk and SRWR on 6 community SSBMs (weak balance). (left) Noise level general to whole graph. (middle) Noise is only inside communities. (right). Noise is only between communities. }
    \label{fig:6com}
\end{figure*}

Here, we perform the same experiments but with a 6 community SSBMs with 50 nodes in each community, and 0.05 probability of link existence between each pair of nodes. As the number of communities increase, the problem of clustering signed networks becomes more difficult. We again perform the semi-supervised clustering with varying levels of noise on the signs of the links.  For the strong and weak walks, we take the state vector at iteration $k = 300$, and set for SRWR the max iterations = 300. The other parameters are chosen by parameters sweep and change for each iteration of the graph. We take the mean and standard deviation over 10 iterations, the results are given in Figure \ref{fig:6com} (left). In this case, we observe that the weak walk performs much better than the strong walk and the SRWR, although the performance decreases quickly as the noise level increases. As in the previous example with 3 communities, we investigate the results when the noise on the signs of the edges is restricted to either inside the communities as in Figure \ref{fig:6com} (centre) or between the communities as in Figure \ref{fig:6com} (right). The weak walk is indeed much more robust to noise inside communities, compared to when the noise is between the communities, outperforming both the strong walk and the SRWR by a larger margin. The distinction between the methods is even more clear than in the previous case of 3 communities. The weak walk, despite having no parameters, outperforms the other methods when the weak balance structure is more clear.

\subsection{Asymmetry}
\begin{figure}[h!]
\centering
\includegraphics[width=0.8\linewidth]{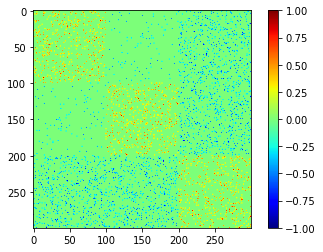}
\caption{Adjacency matrix of 300 node 3-community asymmetric SSBM, with asymmetry parameter $a = 0.2$. }
\label{fig:adj}       
\end{figure}

In most of the previous literature on clustering signed networks, the density of links between each of the communities is held constant. In practice, this may not be the case; different pairs of communities may be connected by different link densities. This would produce, what we call here, an asymmetric SSBM. We define an asymmetry parameter $a$ which controls the link density between communities by defining the link probability matrix as [[0.05, 0.05a, 0.05], [0.05a, 0.05, 0.05], [0.05, 0.05, 0.05]] for a 3 community asymmetric SSBM. One realization of the adjacency matrix for a 3-community asymmetric SSBM with 100 nodes in each community and $a = 0.2$ is given in Figure \ref{fig:adj}. 

\begin{figure}[h!]
\centering
\includegraphics[width=0.8\linewidth]{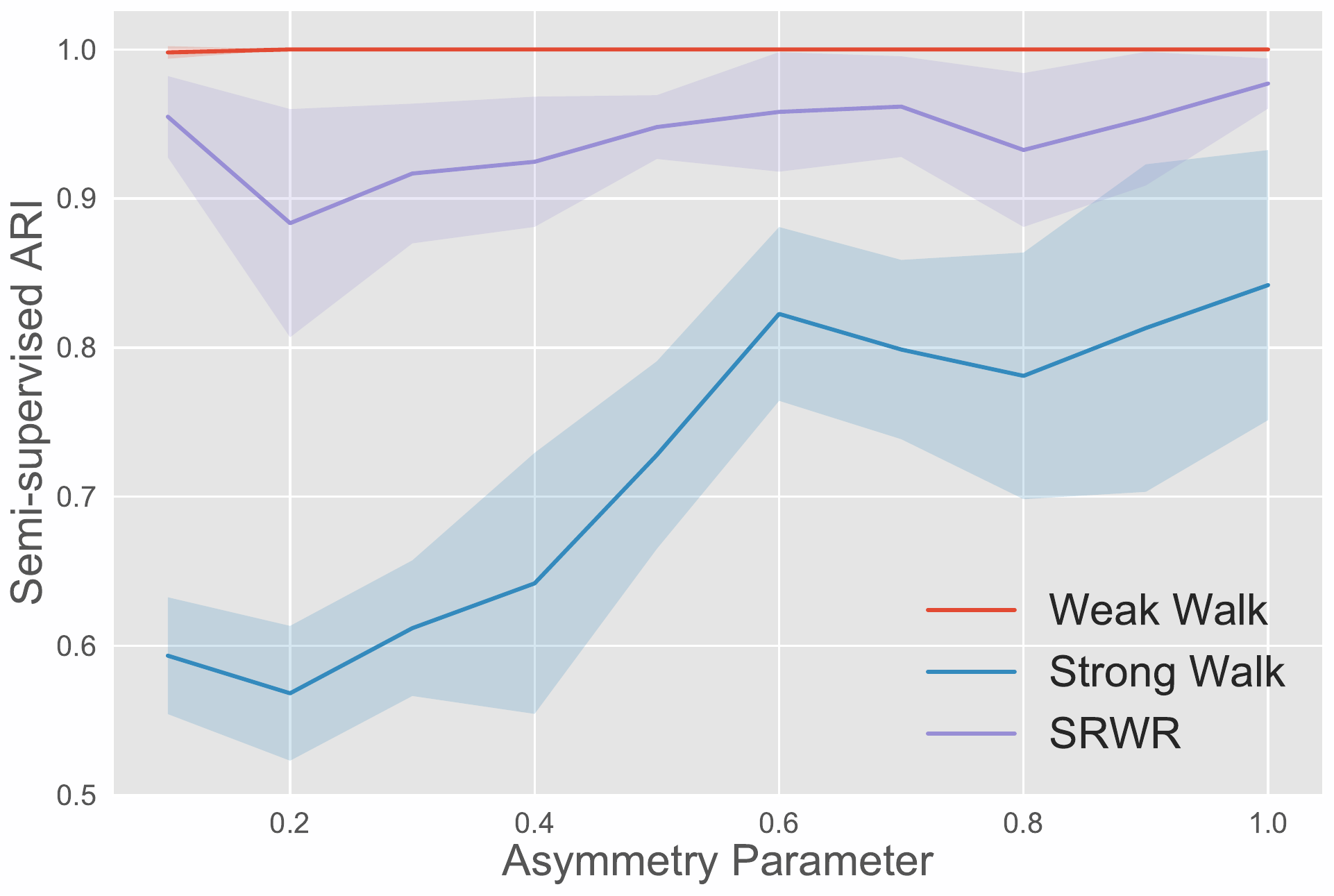}
\caption{ARI from the semi-supervised clustering for the weak walk, strong walk and SRWR on 3 community SSBMs (weak balance), with varying asymmetry. }
\label{fig:asym_3}       
\end{figure}

We perform the semi-supervised clustering on 300 node 3-community SSBMs with varying levels of asymmetry.  For the strong and weak walks, we take the state vector at iteration $k = 100$, and set for SRWR the max iterations = 100. The results are given in Figure \ref{fig:asym_3}. In every cases, the weak walk that we have introduced here outperforms the strong walk and SRWR. This is because walks between the first two communities can happen in two ways, either by taking a negative edge between the two, or by taking two negative edges by passing through the third community. This second walk cannot happen in the case of the weak walk, which cannot take two negative edges. When the asymmetry is high enough, when $a$ is close to 0, the latter case is more likely, and introduces noise into the strong walk, which does not occur in the case of the weak walk. 

\section{Discussion}

Random walks play a central role in the study of networks, and are at the core of several algorithms, e.g. for community detection and network embedding. Random walks allow us to explore the indirect patterns of connectivity between nodes, and their spectral properties are directed related to important network patterns, such as the cut size. The specific rules of a random walk process are a non-trivial choice, however, and it is now recognised that tuning the parameters of the model, e.g. to favour network exploration in depth versus in width, may lead to significant improvements, as was demonstrated by node2vec \cite{grover2016node2vec} for instance. More generally, the choice of a random walk model should be driven by the type of structure that one aims to isolate and/or by the nature of the diffusive processes taking place on an empirical data \cite{lambiotte2019networks}. 

This paper contributes at the problem of designing an ``optimal” random walk model in the case of signed networks. We  argue that the random walk process associated to the standard matrices for the mining of signed networks, e.g. the signed Laplacian, is inherently (and implicitly) biased to the search of a strong balance structure. In the case when the system exhibits multiple clusters, associated to weak  balance, we argue that another random walk model, called the weak walk, is more compatible with the network structure.

After introducing the model and exploring some of its properties, we have considered a classical random-walk based measure of node similarity, in three versions based on the strong walk, the weak walk and the so-called SRWR walk respectively. Through a series of experiments on artificial and real-life data, including unsupervised and semi-supervised clustering, we have shown that the results of weak walks outperform those of strong walks when the graph has more than two communities, or exhibits asymmetry in the density of links. We have also shown that it is competitive against the three-parameter SRWR walk in many cases as well, despite being parameter-free.
  
A wide range of algorithms for network mining take, as an input, trajectories generated by a random walk model. For the design of a similarity matrix, we have made a specific, yet classical, choice for how to combine the multiple walks that may connect a pair of nodes, but other choices could have been made. For example, the similarity matrix could be defined by summing over the walks at each step, by taking the state vector after $k$ steps, taking the stationary state of the walk as in PageRank~\cite{langville2006google}, or by combining the results from walks of different lengths in a non-linear fashion to give greater weight to longer or shorter walks as in the similarity kernel introduced in~\cite{fouss2016algorithms}.
Future research perspectives include feeding weak walk trajectories instead of strong walks ones into other algorithms and  exploring applications  to other typical signed network problems, such as unsupervised clustering \cite{traag2019partitioning} and link prediction \cite{liu2019link}. As a specific example, the strong walk has been used to learn stances for users in signed social networks \cite{pougue2023learning} . Combining weak walks with these state-of-the art methods could  yield, at a minimal developing cost, an improvement in performance for empirical networks, especially when the assumption of strong balance does not hold. 

\section{Methods}
\subsection{Signed Stochastic Block Model}\label{sec:SSBM}
We perform most of our experiments on a signed stochastic block model (SSBM), which is a random graph model with a cluster structure that either exhibits strong or weak balance. The SSBM has $N$ nodes, separated into $k$ clusters. The edge probabilities between each pair of nodes are independent, and are defined by a $k \times k$ probability matrix $P$, where $P_{ij}$ gives the probability that an edge exists between a node in cluster $i$ and another node in cluster $j$. In the initial configuration, edges connecting nodes in the same cluster are positive, while edges connecting nodes in different clusters are negative. As in \cite{cucuringu2019sponge}, unlike in the unsigned case, the edge existence probabilities do not need to be different for inter- and intra-community edges, as the structure of the network is determined by the signs. We introduce a noise parameter called sign flip probability $\nu \in (0,1)$, which gives the probability that the sign of each edge is flipped from the correct sign (positive inside communities and negative between communities). In section VI, we also introduce parameters $\nu_{inside} \in (0,1)$ and $\nu_{between} \in (0,1)$ which separately control the level of noise on the signs of the edges inside and between communities, respectively.

\subsection{Strong/Weak Walk Experiments: Iterative Design}\label{sec:algorithm}
We construct the signed similarity matrix by taking the final state vector at $k = n$, where $n$ is large and the state is approaching the stationary state. We compute the state vector at each step with an iterative process. The way the algorithm defines the similarity matrix is as follows. Beginning with a positive walker on node $x$: 
\begin{equation}
{\bf u}_{0} = {\bf e}_{x}.
\end{equation}
For the strong walk, at each step the new vector ${\bf u}_{k}$ is computed from ${\bf u}_{k-1}$ as: 
\begin{equation}
{\bf u}_{k} = (1-p_{s}) {\bf e}_x + p_{s}{\bf u}_{k-1}T, 
\end{equation}
where $p_{s}$ is the strong walk teleportation parameter. For the weak walk, the process is defined as in Equation \eqref{equ:state-weak}, although it is computationally easier to use the non-Markovian matrix $T_{w}$ to compute ${\bf u}_{k}$ from ${\bf u}_{k-1}$ as: 
\begin{equation}
{\bf u}_{k} =  (1 - (p_w {\bf u}_{i}T_{w})\cdot \boldsymbol{1} 
){\bf e}_{x} + p_w {\bf u}_{k}T_{w}
\label{iterative}
\end{equation}
where $p_{w}$ is the weak walk teleportation parameter, and the first term re-sets the lost probability to the original node $x$ as a positive walker. The advantage of using an interactive process over the matrix given in Equation \eqref{tx} is that this can be done for several initial states at the same time, while the matrix in Equation \eqref{tx} is specific to the state where the walker begins, i.e. node x. However, the processes are equivalent since the term $M_{x}$ in Equation \eqref{tx} returns all probability leakage back to node $x$ as a positive walker, exactly as the first term in Equation \eqref{iterative}.  Finally, for both the strong and weak walks we define the similarity between node $x$ and any other node is given by the vector ${\bf s}_{x}$: 

\begin{equation}
    {\bf s}_{x} = {\bf u}_{k = n}[0:N] - {\bf u}_{k = n}[N+1:2N],
\end{equation}
since we subtract the negative walks from the positive. 

Note that in the experiments we perform, we sample over all possible choices for $p_{s}$, the strong walk teleportation parameter and select the best one for each iteration of the graph. However, we keep $p_w$ fixed at 1, such that the weak walk teleportation rate is determined entirely by the structure of the graph, to keep the weak walk parameter free.

\section{Data Availability}
Most of the experiments are run on synthetic signed networks, which can be generated from the code provided. The data for the Sampson's monastery network is available here: (http://vlado.fmf.uni-lj.si/pub/networks/data/esna/sampson.htm). The data for the Highland Tribes network is available here: (http://konect.cc/networks/ucidata-gama/)

\section{Code Availability}
The code for this study is available at: (https://github.com/saynbabul/weak-walk)

\section{Author Contributions}
All authors conceived the idea, performed the research, and wrote the manuscript.

\section{Competing Interests}

The authors declare no competing interests.

\begin{acknowledgments}
\noindent 
We would like to thank Zoe Schwerkolt for her participation in the early stages of the project, and John Pougu\'e-Biyong for his insightful comments.
R.L. acknowledges support from the EPSRC grants EP/V013068/1, EP/V03474X/1 and EP/Y028872/1. 
S.A.B. was supported by EPSRC grant EP/W523781/1. This work was supported by The Alan Turing Institute’s Enrichment Scheme. Y.T.~is funded by the Wallenberg Initiative on Networks and Quantum Information (WINQ).
\end{acknowledgments}

\bibliography{bib} 

\begin{thebibliography}{44}%
\makeatletter
\providecommand \@ifxundefined [1]{%
 \@ifx{#1\undefined}
}%
\providecommand \@ifnum [1]{%
 \ifnum #1\expandafter \@firstoftwo
 \else \expandafter \@secondoftwo
 \fi
}%
\providecommand \@ifx [1]{%
 \ifx #1\expandafter \@firstoftwo
 \else \expandafter \@secondoftwo
 \fi
}%
\providecommand \natexlab [1]{#1}%
\providecommand \enquote  [1]{``#1''}%
\providecommand \bibnamefont  [1]{#1}%
\providecommand \bibfnamefont [1]{#1}%
\providecommand \citenamefont [1]{#1}%
\providecommand \href@noop [0]{\@secondoftwo}%
\providecommand \href [0]{\begingroup \@sanitize@url \@href}%
\providecommand \@href[1]{\@@startlink{#1}\@@href}%
\providecommand \@@href[1]{\endgroup#1\@@endlink}%
\providecommand \@sanitize@url [0]{\catcode `\\12\catcode `\$12\catcode
  `\&12\catcode `\#12\catcode `\^12\catcode `\_12\catcode `\%12\relax}%
\providecommand \@@startlink[1]{}%
\providecommand \@@endlink[0]{}%
\providecommand \url  [0]{\begingroup\@sanitize@url \@url }%
\providecommand \@url [1]{\endgroup\@href {#1}{\urlprefix }}%
\providecommand \urlprefix  [0]{URL }%
\providecommand \Eprint [0]{\href }%
\providecommand \doibase [0]{http://dx.doi.org/}%
\providecommand \selectlanguage [0]{\@gobble}%
\providecommand \bibinfo  [0]{\@secondoftwo}%
\providecommand \bibfield  [0]{\@secondoftwo}%
\providecommand \translation [1]{[#1]}%
\providecommand \BibitemOpen [0]{}%
\providecommand \bibitemStop [0]{}%
\providecommand \bibitemNoStop [0]{.\EOS\space}%
\providecommand \EOS [0]{\spacefactor3000\relax}%
\providecommand \BibitemShut  [1]{\csname bibitem#1\endcsname}%
\let\auto@bib@innerbib\@empty
\bibitem [{\citenamefont {Masuda}\ \emph {et~al.}(2017)\citenamefont {Masuda},
  \citenamefont {Porter},\ and\ \citenamefont {Lambiotte}}]{Masuda2017RW}%
  \BibitemOpen
  \bibfield  {author} {\bibinfo {author} {\bibfnamefont {N.}~\bibnamefont
  {Masuda}}, \bibinfo {author} {\bibfnamefont {M.}~\bibnamefont {Porter}}, \
  and\ \bibinfo {author} {\bibfnamefont {R.}~\bibnamefont {Lambiotte}},\ }\href
  {\doibase 10.1016/j.physrep.2017.07.007} {\bibfield  {journal} {\bibinfo
  {journal} {Phys. Rep.}\ }\textbf {\bibinfo {volume} {716--717}},\ \bibinfo
  {pages} {1} (\bibinfo {year} {2017})}\BibitemShut {NoStop}%
\bibitem [{\citenamefont {Lambiotte}\ and\ \citenamefont
  {Schaub}(2021)}]{lambiotte2021modularity}%
  \BibitemOpen
  \bibfield  {author} {\bibinfo {author} {\bibfnamefont {R.}~\bibnamefont
  {Lambiotte}}\ and\ \bibinfo {author} {\bibfnamefont {M.~T.}\ \bibnamefont
  {Schaub}},\ }\href@noop {} {\emph {\bibinfo {title} {Modularity and dynamics
  on complex networks}}}\ (\bibinfo  {publisher} {Cambridge University Press},\
  \bibinfo {year} {2021})\BibitemShut {NoStop}%
\bibitem [{\citenamefont {Levin}\ and\ \citenamefont
  {Peres}(2017)}]{levin2017markov}%
  \BibitemOpen
  \bibfield  {author} {\bibinfo {author} {\bibfnamefont {D.~A.}\ \bibnamefont
  {Levin}}\ and\ \bibinfo {author} {\bibfnamefont {Y.}~\bibnamefont {Peres}},\
  }\href@noop {} {\emph {\bibinfo {title} {Markov chains and mixing times}}},\
  Vol.\ \bibinfo {volume} {107}\ (\bibinfo  {publisher} {American Mathematical
  Soc.},\ \bibinfo {year} {2017})\BibitemShut {NoStop}%
\bibitem [{\citenamefont {Gleich}(2015)}]{gleich2015pagerank}%
  \BibitemOpen
  \bibfield  {author} {\bibinfo {author} {\bibfnamefont {D.~F.}\ \bibnamefont
  {Gleich}},\ }\href@noop {} {\bibfield  {journal} {\bibinfo  {journal} {siam
  REVIEW}\ }\textbf {\bibinfo {volume} {57}},\ \bibinfo {pages} {321} (\bibinfo
  {year} {2015})}\BibitemShut {NoStop}%
\bibitem [{\citenamefont {Rosvall}\ and\ \citenamefont
  {Bergstrom}(2008)}]{rosvall2008maps}%
  \BibitemOpen
  \bibfield  {author} {\bibinfo {author} {\bibfnamefont {M.}~\bibnamefont
  {Rosvall}}\ and\ \bibinfo {author} {\bibfnamefont {C.~T.}\ \bibnamefont
  {Bergstrom}},\ }\href@noop {} {\bibfield  {journal} {\bibinfo  {journal}
  {Proceedings of the national academy of sciences}\ }\textbf {\bibinfo
  {volume} {105}},\ \bibinfo {pages} {1118} (\bibinfo {year}
  {2008})}\BibitemShut {NoStop}%
\bibitem [{\citenamefont {Lambiotte}\ \emph
  {et~al.}(2014{\natexlab{a}})\citenamefont {Lambiotte}, \citenamefont
  {Delvenne},\ and\ \citenamefont {Barahona}}]{lambiotte2014random}%
  \BibitemOpen
  \bibfield  {author} {\bibinfo {author} {\bibfnamefont {R.}~\bibnamefont
  {Lambiotte}}, \bibinfo {author} {\bibfnamefont {J.-C.}\ \bibnamefont
  {Delvenne}}, \ and\ \bibinfo {author} {\bibfnamefont {M.}~\bibnamefont
  {Barahona}},\ }\href@noop {} {\bibfield  {journal} {\bibinfo  {journal} {IEEE
  Transactions on Network Science and Engineering}\ }\textbf {\bibinfo {volume}
  {1}},\ \bibinfo {pages} {76} (\bibinfo {year}
  {2014}{\natexlab{a}})}\BibitemShut {NoStop}%
\bibitem [{\citenamefont {Fouss}\ \emph {et~al.}(2016)\citenamefont {Fouss},
  \citenamefont {Saerens},\ and\ \citenamefont {Shimbo}}]{fouss2016algorithms}%
  \BibitemOpen
  \bibfield  {author} {\bibinfo {author} {\bibfnamefont {F.}~\bibnamefont
  {Fouss}}, \bibinfo {author} {\bibfnamefont {M.}~\bibnamefont {Saerens}}, \
  and\ \bibinfo {author} {\bibfnamefont {M.}~\bibnamefont {Shimbo}},\
  }\href@noop {} {\emph {\bibinfo {title} {Algorithms and models for network
  data and link analysis}}}\ (\bibinfo  {publisher} {Cambridge University
  Press},\ \bibinfo {year} {2016})\BibitemShut {NoStop}%
\bibitem [{\citenamefont {Grover}\ and\ \citenamefont
  {Leskovec}(2016)}]{grover2016node2vec}%
  \BibitemOpen
  \bibfield  {author} {\bibinfo {author} {\bibfnamefont {A.}~\bibnamefont
  {Grover}}\ and\ \bibinfo {author} {\bibfnamefont {J.}~\bibnamefont
  {Leskovec}},\ }in\ \href@noop {} {\emph {\bibinfo {booktitle} {Proceedings of
  the 22nd ACM SIGKDD international conference on Knowledge discovery and data
  mining}}}\ (\bibinfo {year} {2016})\ pp.\ \bibinfo {pages}
  {855--864}\BibitemShut {NoStop}%
\bibitem [{\citenamefont {Altafini}(2012)}]{altafini2012dynamics}%
  \BibitemOpen
  \bibfield  {author} {\bibinfo {author} {\bibfnamefont {C.}~\bibnamefont
  {Altafini}},\ }\href@noop {} {\bibfield  {journal} {\bibinfo  {journal} {PloS
  one}\ }\textbf {\bibinfo {volume} {7}},\ \bibinfo {pages} {e38135} (\bibinfo
  {year} {2012})}\BibitemShut {NoStop}%
\bibitem [{\citenamefont {Traag}\ \emph {et~al.}(2019)\citenamefont {Traag},
  \citenamefont {Doreian},\ and\ \citenamefont
  {Mrvar}}]{traag2019partitioning}%
  \BibitemOpen
  \bibfield  {author} {\bibinfo {author} {\bibfnamefont {V.}~\bibnamefont
  {Traag}}, \bibinfo {author} {\bibfnamefont {P.}~\bibnamefont {Doreian}}, \
  and\ \bibinfo {author} {\bibfnamefont {A.}~\bibnamefont {Mrvar}},\
  }\href@noop {} {\bibfield  {journal} {\bibinfo  {journal} {Advances in
  network clustering and blockmodeling}\ ,\ \bibinfo {pages} {225}} (\bibinfo
  {year} {2019})}\BibitemShut {NoStop}%
\bibitem [{\citenamefont {Szell}\ \emph {et~al.}(2010)\citenamefont {Szell},
  \citenamefont {Lambiotte},\ and\ \citenamefont
  {Thurner}}]{szell2010multirelational}%
  \BibitemOpen
  \bibfield  {author} {\bibinfo {author} {\bibfnamefont {M.}~\bibnamefont
  {Szell}}, \bibinfo {author} {\bibfnamefont {R.}~\bibnamefont {Lambiotte}}, \
  and\ \bibinfo {author} {\bibfnamefont {S.}~\bibnamefont {Thurner}},\
  }\href@noop {} {\bibfield  {journal} {\bibinfo  {journal} {Proceedings of the
  National Academy of Sciences}\ }\textbf {\bibinfo {volume} {107}},\ \bibinfo
  {pages} {13636} (\bibinfo {year} {2010})}\BibitemShut {NoStop}%
\bibitem [{\citenamefont {Kunegis}\ \emph {et~al.}(2010)\citenamefont
  {Kunegis}, \citenamefont {Schmidt}, \citenamefont {Lommatzsch}, \citenamefont
  {Lerner}, \citenamefont {De~Luca},\ and\ \citenamefont
  {Albayrak}}]{kunegis2010spectral}%
  \BibitemOpen
  \bibfield  {author} {\bibinfo {author} {\bibfnamefont {J.}~\bibnamefont
  {Kunegis}}, \bibinfo {author} {\bibfnamefont {S.}~\bibnamefont {Schmidt}},
  \bibinfo {author} {\bibfnamefont {A.}~\bibnamefont {Lommatzsch}}, \bibinfo
  {author} {\bibfnamefont {J.}~\bibnamefont {Lerner}}, \bibinfo {author}
  {\bibfnamefont {E.~W.}\ \bibnamefont {De~Luca}}, \ and\ \bibinfo {author}
  {\bibfnamefont {S.}~\bibnamefont {Albayrak}},\ }in\ \href@noop {} {\emph
  {\bibinfo {booktitle} {Proceedings of the 2010 SIAM international conference
  on data mining}}}\ (\bibinfo {organization} {SIAM},\ \bibinfo {year} {2010})\
  pp.\ \bibinfo {pages} {559--570}\BibitemShut {NoStop}%
\bibitem [{\citenamefont {Tian}\ and\ \citenamefont
  {Lambiotte}(2024)}]{tian2024spreading}%
  \BibitemOpen
  \bibfield  {author} {\bibinfo {author} {\bibfnamefont {Y.}~\bibnamefont
  {Tian}}\ and\ \bibinfo {author} {\bibfnamefont {R.}~\bibnamefont
  {Lambiotte}},\ }\href@noop {} {\bibfield  {journal} {\bibinfo  {journal}
  {SIAM Journal on Applied Dynamical Systems}\ }\textbf {\bibinfo {volume}
  {23}},\ \bibinfo {pages} {50} (\bibinfo {year} {2024})}\BibitemShut {NoStop}%
\bibitem [{\citenamefont {Cheeger}(1970)}]{cheeger1970lower}%
  \BibitemOpen
  \bibfield  {author} {\bibinfo {author} {\bibfnamefont {J.}~\bibnamefont
  {Cheeger}},\ }\href@noop {} {\bibfield  {journal} {\bibinfo  {journal}
  {Problems in analysis}\ }\textbf {\bibinfo {volume} {625}},\ \bibinfo {pages}
  {110} (\bibinfo {year} {1970})}\BibitemShut {NoStop}%
\bibitem [{\citenamefont {Harary}(1953)}]{harary1953notion}%
  \BibitemOpen
  \bibfield  {author} {\bibinfo {author} {\bibfnamefont {F.}~\bibnamefont
  {Harary}},\ }\href@noop {} {\bibfield  {journal} {\bibinfo  {journal}
  {Michigan Mathematical Journal}\ }\textbf {\bibinfo {volume} {2}},\ \bibinfo
  {pages} {143} (\bibinfo {year} {1953})}\BibitemShut {NoStop}%
\bibitem [{\citenamefont {Davis}(1967)}]{davis1967clustering}%
  \BibitemOpen
  \bibfield  {author} {\bibinfo {author} {\bibfnamefont {J.~A.}\ \bibnamefont
  {Davis}},\ }\href@noop {} {\bibfield  {journal} {\bibinfo  {journal} {Human
  relations}\ }\textbf {\bibinfo {volume} {20}},\ \bibinfo {pages} {181}
  (\bibinfo {year} {1967})}\BibitemShut {NoStop}%
\bibitem [{\citenamefont {Cucuringu}\ \emph {et~al.}(2019)\citenamefont
  {Cucuringu}, \citenamefont {Davies}, \citenamefont {Glielmo},\ and\
  \citenamefont {Tyagi}}]{cucuringu2019sponge}%
  \BibitemOpen
  \bibfield  {author} {\bibinfo {author} {\bibfnamefont {M.}~\bibnamefont
  {Cucuringu}}, \bibinfo {author} {\bibfnamefont {P.}~\bibnamefont {Davies}},
  \bibinfo {author} {\bibfnamefont {A.}~\bibnamefont {Glielmo}}, \ and\
  \bibinfo {author} {\bibfnamefont {H.}~\bibnamefont {Tyagi}},\ }in\ \href@noop
  {} {\emph {\bibinfo {booktitle} {The 22nd International Conference on
  Artificial Intelligence and Statistics}}}\ (\bibinfo {organization} {PMLR},\
  \bibinfo {year} {2019})\ pp.\ \bibinfo {pages} {1088--1098}\BibitemShut
  {NoStop}%
\bibitem [{\citenamefont {Babul}\ and\ \citenamefont
  {Lambiotte}(2024)}]{babul2024sheep}%
  \BibitemOpen
  \bibfield  {author} {\bibinfo {author} {\bibfnamefont {S.}~\bibnamefont
  {Babul}}\ and\ \bibinfo {author} {\bibfnamefont {R.}~\bibnamefont
  {Lambiotte}},\ }\href@noop {} {\bibfield  {journal} {\bibinfo  {journal}
  {Communications Physics}\ }\textbf {\bibinfo {volume} {7}},\ \bibinfo {pages}
  {8} (\bibinfo {year} {2024})}\BibitemShut {NoStop}%
\bibitem [{\citenamefont {Perozzi}\ \emph {et~al.}(2014)\citenamefont
  {Perozzi}, \citenamefont {Al-Rfou},\ and\ \citenamefont
  {Skiena}}]{perozzi2014deepwalk}%
  \BibitemOpen
  \bibfield  {author} {\bibinfo {author} {\bibfnamefont {B.}~\bibnamefont
  {Perozzi}}, \bibinfo {author} {\bibfnamefont {R.}~\bibnamefont {Al-Rfou}}, \
  and\ \bibinfo {author} {\bibfnamefont {S.}~\bibnamefont {Skiena}},\ }in\
  \href@noop {} {\emph {\bibinfo {booktitle} {Proceedings of the 20th ACM
  SIGKDD international conference on Knowledge discovery and data mining}}}\
  (\bibinfo {year} {2014})\ pp.\ \bibinfo {pages} {701--710}\BibitemShut
  {NoStop}%
\bibitem [{Note1()}]{Note1}%
  \BibitemOpen
  \bibinfo {note} {This is the case in the vast majority of real-world
  applications, especially when the graph is sufficiently large.}\BibitemShut
  {Stop}%
\bibitem [{\citenamefont {ans J.~Lafferty}(2002)}]{kondor2002DiffusionKO}%
  \BibitemOpen
  \bibfield  {author} {\bibinfo {author} {\bibfnamefont {R.~K.}\ \bibnamefont
  {ans J.~Lafferty}},\ }in\ \href@noop {} {\emph {\bibinfo {booktitle}
  {International Conference on Machine Learning}}}\ (\bibinfo {year} {2002})\
  p.\ \bibinfo {pages} {315–322}\BibitemShut {NoStop}%
\bibitem [{\citenamefont {Lawler}(2010)}]{lawler2010heat}%
  \BibitemOpen
  \bibfield  {author} {\bibinfo {author} {\bibfnamefont {G.}~\bibnamefont
  {Lawler}},\ }\href@noop {} {\emph {\bibinfo {title} {Random walk and the heat
  equation}}}\ (\bibinfo  {publisher} {American Mathematical Society},\
  \bibinfo {year} {2010})\BibitemShut {NoStop}%
\bibitem [{\citenamefont {Smola}\ and\ \citenamefont
  {Kondor}(2003)}]{smola2003heat}%
  \BibitemOpen
  \bibfield  {author} {\bibinfo {author} {\bibfnamefont {A.}~\bibnamefont
  {Smola}}\ and\ \bibinfo {author} {\bibfnamefont {R.}~\bibnamefont {Kondor}},\
  }in\ \href@noop {} {\emph {\bibinfo {booktitle} {Learning Theory and Kernel
  Machines}}},\ \bibinfo {editor} {edited by\ \bibinfo {editor} {\bibfnamefont
  {B.}~\bibnamefont {Sch{\"o}lkopf}}\ and\ \bibinfo {editor} {\bibfnamefont
  {M.}~\bibnamefont {Warmuth}}}\ (\bibinfo  {publisher} {Springer Berlin
  Heidelberg},\ \bibinfo {address} {Berlin, Heidelberg},\ \bibinfo {year}
  {2003})\ pp.\ \bibinfo {pages} {144--158}\BibitemShut {NoStop}%
\bibitem [{\citenamefont {Fouss}\ \emph {et~al.}(2007)\citenamefont {Fouss},
  \citenamefont {Pirotte}, \citenamefont {Renders},\ and\ \citenamefont
  {Saerens}}]{fouss2007communte}%
  \BibitemOpen
  \bibfield  {author} {\bibinfo {author} {\bibfnamefont {F.}~\bibnamefont
  {Fouss}}, \bibinfo {author} {\bibfnamefont {A.}~\bibnamefont {Pirotte}},
  \bibinfo {author} {\bibfnamefont {J.}~\bibnamefont {Renders}}, \ and\
  \bibinfo {author} {\bibfnamefont {M.}~\bibnamefont {Saerens}},\ }\href
  {\doibase 10.1109/TKDE.2007.46} {\bibfield  {journal} {\bibinfo  {journal}
  {IEEE Transactions on Knowledge and Data Engineering}\ }\textbf {\bibinfo
  {volume} {19}},\ \bibinfo {pages} {355} (\bibinfo {year} {2007})}\BibitemShut
  {NoStop}%
\bibitem [{\citenamefont {Saerens}\ \emph {et~al.}(2004)\citenamefont
  {Saerens}, \citenamefont {Fouss}, \citenamefont {Yen},\ and\ \citenamefont
  {Dupont}}]{saerens2004commute}%
  \BibitemOpen
  \bibfield  {author} {\bibinfo {author} {\bibfnamefont {M.}~\bibnamefont
  {Saerens}}, \bibinfo {author} {\bibfnamefont {F.}~\bibnamefont {Fouss}},
  \bibinfo {author} {\bibfnamefont {L.}~\bibnamefont {Yen}}, \ and\ \bibinfo
  {author} {\bibfnamefont {P.}~\bibnamefont {Dupont}},\ }in\ \href@noop {}
  {\emph {\bibinfo {booktitle} {Machine Learning: ECML 2004}}},\ \bibinfo
  {editor} {edited by\ \bibinfo {editor} {\bibfnamefont {J.}~\bibnamefont
  {Boulicaut}}, \bibinfo {editor} {\bibfnamefont {F.}~\bibnamefont {Esposito}},
  \bibinfo {editor} {\bibfnamefont {F.}~\bibnamefont {Giannotti}}, \ and\
  \bibinfo {editor} {\bibfnamefont {D.}~\bibnamefont {Pedreschi}}}\ (\bibinfo
  {publisher} {Springer Berlin Heidelberg},\ \bibinfo {address} {Berlin,
  Heidelberg},\ \bibinfo {year} {2004})\ pp.\ \bibinfo {pages}
  {371--383}\BibitemShut {NoStop}%
\bibitem [{\citenamefont {Pan}\ \emph {et~al.}(2004)\citenamefont {Pan},
  \citenamefont {Yang}, \citenamefont {Faloutsos},\ and\ \citenamefont
  {Duygulu}}]{pan2004rwr}%
  \BibitemOpen
  \bibfield  {author} {\bibinfo {author} {\bibfnamefont {J.}~\bibnamefont
  {Pan}}, \bibinfo {author} {\bibfnamefont {H.}~\bibnamefont {Yang}}, \bibinfo
  {author} {\bibfnamefont {C.}~\bibnamefont {Faloutsos}}, \ and\ \bibinfo
  {author} {\bibfnamefont {P.}~\bibnamefont {Duygulu}},\ }in\ \href {\doibase
  10.1145/1014052.1014135} {\emph {\bibinfo {booktitle} {Proceedings of the
  Tenth ACM SIGKDD International Conference on Knowledge Discovery and Data
  Mining}}}\ (\bibinfo  {publisher} {Association for Computing Machinery},\
  \bibinfo {address} {New York, NY, USA},\ \bibinfo {year} {2004})\ pp.\
  \bibinfo {pages} {653--658}\BibitemShut {NoStop}%
\bibitem [{\citenamefont {Tong}\ \emph {et~al.}(2007)\citenamefont {Tong},
  \citenamefont {Faloutsos},\ and\ \citenamefont {Koren}}]{tong2007rwr}%
  \BibitemOpen
  \bibfield  {author} {\bibinfo {author} {\bibfnamefont {H.}~\bibnamefont
  {Tong}}, \bibinfo {author} {\bibfnamefont {C.}~\bibnamefont {Faloutsos}}, \
  and\ \bibinfo {author} {\bibfnamefont {Y.}~\bibnamefont {Koren}},\ }in\ \href
  {\doibase 10.1145/1281192.1281272} {\emph {\bibinfo {booktitle} {Proceedings
  of the 13th ACM SIGKDD International Conference on Knowledge Discovery and
  Data Mining}}}\ (\bibinfo  {publisher} {Association for Computing
  Machinery},\ \bibinfo {address} {New York, NY, USA},\ \bibinfo {year}
  {2007})\ pp.\ \bibinfo {pages} {747--756}\BibitemShut {NoStop}%
\bibitem [{\citenamefont {Tong}\ \emph {et~al.}(2008)\citenamefont {Tong},
  \citenamefont {Faloutsos},\ and\ \citenamefont {Pan}}]{tong2008rwr}%
  \BibitemOpen
  \bibfield  {author} {\bibinfo {author} {\bibfnamefont {H.}~\bibnamefont
  {Tong}}, \bibinfo {author} {\bibfnamefont {C.}~\bibnamefont {Faloutsos}}, \
  and\ \bibinfo {author} {\bibfnamefont {J.-Y.}\ \bibnamefont {Pan}},\
  }\href@noop {} {\bibfield  {journal} {\bibinfo  {journal} {Knowl. Inf.
  Syst.}\ }\textbf {\bibinfo {volume} {14}},\ \bibinfo {pages} {327} (\bibinfo
  {year} {2008})}\BibitemShut {NoStop}%
\bibitem [{\citenamefont {Qiu}\ \emph {et~al.}(2018)\citenamefont {Qiu},
  \citenamefont {Dong}, \citenamefont {Ma}, \citenamefont {Li}, \citenamefont
  {Wang},\ and\ \citenamefont {Tang}}]{qiu2018network}%
  \BibitemOpen
  \bibfield  {author} {\bibinfo {author} {\bibfnamefont {J.}~\bibnamefont
  {Qiu}}, \bibinfo {author} {\bibfnamefont {Y.}~\bibnamefont {Dong}}, \bibinfo
  {author} {\bibfnamefont {H.}~\bibnamefont {Ma}}, \bibinfo {author}
  {\bibfnamefont {J.}~\bibnamefont {Li}}, \bibinfo {author} {\bibfnamefont
  {K.}~\bibnamefont {Wang}}, \ and\ \bibinfo {author} {\bibfnamefont
  {J.}~\bibnamefont {Tang}},\ }in\ \href@noop {} {\emph {\bibinfo {booktitle}
  {Proceedings of the eleventh ACM international conference on web search and
  data mining}}}\ (\bibinfo {year} {2018})\ pp.\ \bibinfo {pages}
  {459--467}\BibitemShut {NoStop}%
\bibitem [{Note2()}]{Note2}%
  \BibitemOpen
  \bibinfo {note} {The symmetry arises due to the detailed balance when
  teleportation to nodes is done proportional to the stationary state of the
  walk without restart.}\BibitemShut {Stop}%
\bibitem [{\citenamefont {Girvan}\ and\ \citenamefont
  {Newman}(2002)}]{girvan2002modularity}%
  \BibitemOpen
  \bibfield  {author} {\bibinfo {author} {\bibfnamefont {M.}~\bibnamefont
  {Girvan}}\ and\ \bibinfo {author} {\bibfnamefont {M.}~\bibnamefont
  {Newman}},\ }\href {\doibase 10.1073/pnas.122653799} {\bibfield  {journal}
  {\bibinfo  {journal} {Proceedings of the National Academy of Sciences}\
  }\textbf {\bibinfo {volume} {99}},\ \bibinfo {pages} {7821} (\bibinfo {year}
  {2002})}\BibitemShut {NoStop}%
\bibitem [{\citenamefont {Delvenne}\ \emph {et~al.}(2010)\citenamefont
  {Delvenne}, \citenamefont {Yaliraki},\ and\ \citenamefont
  {Barahona}}]{delvenne2010markovs}%
  \BibitemOpen
  \bibfield  {author} {\bibinfo {author} {\bibfnamefont {J.}~\bibnamefont
  {Delvenne}}, \bibinfo {author} {\bibfnamefont {S.}~\bibnamefont {Yaliraki}},
  \ and\ \bibinfo {author} {\bibfnamefont {M.}~\bibnamefont {Barahona}},\
  }\href {\doibase 10.1073/pnas.0903215107} {\bibfield  {journal} {\bibinfo
  {journal} {Proceedings of the National Academy of Sciences}\ }\textbf
  {\bibinfo {volume} {107}},\ \bibinfo {pages} {12755} (\bibinfo {year}
  {2010})}\BibitemShut {NoStop}%
\bibitem [{\citenamefont {Lambiotte}\ \emph
  {et~al.}(2014{\natexlab{b}})\citenamefont {Lambiotte}, \citenamefont
  {Delvenne},\ and\ \citenamefont {Barahona}}]{lambiotte2014rws}%
  \BibitemOpen
  \bibfield  {author} {\bibinfo {author} {\bibfnamefont {R.}~\bibnamefont
  {Lambiotte}}, \bibinfo {author} {\bibfnamefont {J.}~\bibnamefont {Delvenne}},
  \ and\ \bibinfo {author} {\bibfnamefont {M.}~\bibnamefont {Barahona}},\
  }\href@noop {} {\bibfield  {journal} {\bibinfo  {journal} {IEEE Transactions
  on Network Science and Engineering}\ }\textbf {\bibinfo {volume} {1}},\
  \bibinfo {pages} {76} (\bibinfo {year} {2014}{\natexlab{b}})}\BibitemShut
  {NoStop}%
\bibitem [{\citenamefont {Schaub}\ \emph {et~al.}(2012)\citenamefont {Schaub},
  \citenamefont {Delvenne}, \citenamefont {Yaliraki},\ and\ \citenamefont
  {Barahona}}]{schaub2012rws}%
  \BibitemOpen
  \bibfield  {author} {\bibinfo {author} {\bibfnamefont {M.}~\bibnamefont
  {Schaub}}, \bibinfo {author} {\bibfnamefont {J.}~\bibnamefont {Delvenne}},
  \bibinfo {author} {\bibfnamefont {S.}~\bibnamefont {Yaliraki}}, \ and\
  \bibinfo {author} {\bibfnamefont {M.}~\bibnamefont {Barahona}},\ }\href
  {\doibase 10.1371/journal.pone.0032210} {\bibfield  {journal} {\bibinfo
  {journal} {PLOS ONE}\ }\textbf {\bibinfo {volume} {7}},\ \bibinfo {pages} {1}
  (\bibinfo {year} {2012})}\BibitemShut {NoStop}%
\bibitem [{\citenamefont {Jung}\ \emph {et~al.}(2016)\citenamefont {Jung},
  \citenamefont {Jin}, \citenamefont {Sael},\ and\ \citenamefont
  {Kang}}]{7837935}%
  \BibitemOpen
  \bibfield  {author} {\bibinfo {author} {\bibfnamefont {J.}~\bibnamefont
  {Jung}}, \bibinfo {author} {\bibfnamefont {W.}~\bibnamefont {Jin}}, \bibinfo
  {author} {\bibfnamefont {L.}~\bibnamefont {Sael}}, \ and\ \bibinfo {author}
  {\bibfnamefont {U.}~\bibnamefont {Kang}},\ }in\ \href {\doibase
  10.1109/ICDM.2016.0122} {\emph {\bibinfo {booktitle} {2016 IEEE 16th
  International Conference on Data Mining (ICDM)}}}\ (\bibinfo {year} {2016})\
  pp.\ \bibinfo {pages} {973--978}\BibitemShut {NoStop}%
\bibitem [{\citenamefont {Langville}\ and\ \citenamefont
  {Meyer}(2006)}]{langville2006google}%
  \BibitemOpen
  \bibfield  {author} {\bibinfo {author} {\bibfnamefont {A.~N.}\ \bibnamefont
  {Langville}}\ and\ \bibinfo {author} {\bibfnamefont {C.~D.}\ \bibnamefont
  {Meyer}},\ }\href@noop {} {\emph {\bibinfo {title} {Google's PageRank and
  beyond: The science of search engine rankings}}}\ (\bibinfo  {publisher}
  {Princeton university press},\ \bibinfo {year} {2006})\BibitemShut {NoStop}%
\bibitem [{\citenamefont {Gates}\ and\ \citenamefont
  {Ahn}(2017)}]{gates2017impact}%
  \BibitemOpen
  \bibfield  {author} {\bibinfo {author} {\bibfnamefont {A.~J.}\ \bibnamefont
  {Gates}}\ and\ \bibinfo {author} {\bibfnamefont {Y.-Y.}\ \bibnamefont
  {Ahn}},\ }\href@noop {} {\bibfield  {journal} {\bibinfo  {journal} {Journal
  of Machine Learning Research}\ }\textbf {\bibinfo {volume} {18}},\ \bibinfo
  {pages} {1} (\bibinfo {year} {2017})}\BibitemShut {NoStop}%
\bibitem [{\citenamefont {Read}(1954)}]{read1954cultures}%
  \BibitemOpen
  \bibfield  {author} {\bibinfo {author} {\bibfnamefont {K.~E.}\ \bibnamefont
  {Read}},\ }\href@noop {} {\bibfield  {journal} {\bibinfo  {journal}
  {Southwestern Journal of Anthropology}\ ,\ \bibinfo {pages} {1}} (\bibinfo
  {year} {1954})}\BibitemShut {NoStop}%
\bibitem [{\citenamefont {Hage}\ and\ \citenamefont
  {Harary}(1983)}]{hage1983structural}%
  \BibitemOpen
  \bibfield  {author} {\bibinfo {author} {\bibfnamefont {P.}~\bibnamefont
  {Hage}}\ and\ \bibinfo {author} {\bibfnamefont {F.}~\bibnamefont {Harary}},\
  }\href@noop {} {\bibfield  {journal} {\bibinfo  {journal} {(No Title)}\ }
  (\bibinfo {year} {1983})}\BibitemShut {NoStop}%
\bibitem [{\citenamefont {Traag}\ and\ \citenamefont
  {Bruggeman}(2009)}]{traag2009community}%
  \BibitemOpen
  \bibfield  {author} {\bibinfo {author} {\bibfnamefont {V.~A.}\ \bibnamefont
  {Traag}}\ and\ \bibinfo {author} {\bibfnamefont {J.}~\bibnamefont
  {Bruggeman}},\ }\href@noop {} {\bibfield  {journal} {\bibinfo  {journal}
  {Physical Review E}\ }\textbf {\bibinfo {volume} {80}},\ \bibinfo {pages}
  {036115} (\bibinfo {year} {2009})}\BibitemShut {NoStop}%
\bibitem [{\citenamefont {Sampson}(1968)}]{sampson1968novitiate}%
  \BibitemOpen
  \bibfield  {author} {\bibinfo {author} {\bibfnamefont {S.~F.}\ \bibnamefont
  {Sampson}},\ }\href@noop {} {\emph {\bibinfo {title} {A novitiate in a period
  of change: An experimental and case study of social relationships}}}\
  (\bibinfo  {publisher} {Cornell University},\ \bibinfo {year}
  {1968})\BibitemShut {NoStop}%
\bibitem [{\citenamefont {Lambiotte}\ \emph {et~al.}(2019)\citenamefont
  {Lambiotte}, \citenamefont {Rosvall},\ and\ \citenamefont
  {Scholtes}}]{lambiotte2019networks}%
  \BibitemOpen
  \bibfield  {author} {\bibinfo {author} {\bibfnamefont {R.}~\bibnamefont
  {Lambiotte}}, \bibinfo {author} {\bibfnamefont {M.}~\bibnamefont {Rosvall}},
  \ and\ \bibinfo {author} {\bibfnamefont {I.}~\bibnamefont {Scholtes}},\
  }\href@noop {} {\bibfield  {journal} {\bibinfo  {journal} {Nature physics}\
  }\textbf {\bibinfo {volume} {15}},\ \bibinfo {pages} {313} (\bibinfo {year}
  {2019})}\BibitemShut {NoStop}%
\bibitem [{\citenamefont {Liu}\ \emph {et~al.}(2019)\citenamefont {Liu},
  \citenamefont {Xiao},\ and\ \citenamefont {Xu}}]{liu2019link}%
  \BibitemOpen
  \bibfield  {author} {\bibinfo {author} {\bibfnamefont {S.-Y.}\ \bibnamefont
  {Liu}}, \bibinfo {author} {\bibfnamefont {J.}~\bibnamefont {Xiao}}, \ and\
  \bibinfo {author} {\bibfnamefont {X.-K.}\ \bibnamefont {Xu}},\ }\href@noop {}
  {\bibfield  {journal} {\bibinfo  {journal} {IEEE Transactions on Network
  Science and Engineering}\ }\textbf {\bibinfo {volume} {7}},\ \bibinfo {pages}
  {1724} (\bibinfo {year} {2019})}\BibitemShut {NoStop}%
\bibitem [{\citenamefont {Pougu{\'e}-Biyong}\ \emph {et~al.}(2023)\citenamefont
  {Pougu{\'e}-Biyong}, \citenamefont {Gupta}, \citenamefont {Haghighi},\ and\
  \citenamefont {El-Kishky}}]{pougue2023learning}%
  \BibitemOpen
  \bibfield  {author} {\bibinfo {author} {\bibfnamefont {J.}~\bibnamefont
  {Pougu{\'e}-Biyong}}, \bibinfo {author} {\bibfnamefont {A.}~\bibnamefont
  {Gupta}}, \bibinfo {author} {\bibfnamefont {A.}~\bibnamefont {Haghighi}}, \
  and\ \bibinfo {author} {\bibfnamefont {A.}~\bibnamefont {El-Kishky}},\ }in\
  \href@noop {} {\emph {\bibinfo {booktitle} {Proceedings of the Sixteenth ACM
  International Conference on Web Search and Data Mining}}}\ (\bibinfo {year}
  {2023})\ pp.\ \bibinfo {pages} {177--185}\BibitemShut {NoStop}%
\end{thebibliography}%

\end{document}